\documentclass[12pt]{article}
\usepackage{xr-hyper}

\usepackage{xr}
\usepackage{amssymb}
\usepackage{amsfonts}
\usepackage{amsmath}
\usepackage{theorem}
\usepackage{graphicx}
\usepackage{xcolor}
\usepackage[longnamesfirst]{natbib}

\newtheorem{assumption}{Assumption}[section]
\newtheorem{theorem}{Theorem}[section]

\newtheorem{algorithm}[theorem]{Algorithm}

\newtheorem{definition}[theorem]{Definition}

\newtheorem{remark}{Remark}[section]

\newenvironment{proof}[1][Proof]{\textbf{#1.} }{\ \rule{0.5em}{0.5em}}

\allowdisplaybreaks
\usepackage{multirow}
\usepackage{rotating}
\usepackage{caption}
\usepackage{enumitem}
\usepackage[hidelinks]{hyperref}

\newcommand{\E}{\mathbb{E}}
\newcommand{\Var}{\text{Var}}
\renewcommand{\P}{\mathbb{P}}
\newcommand{\I}{\mathcal{I}}

\newcommand{\1}{\mathbf{1}}

\newcommand{\sumi}{\sum_{i=1}^N}

\newcommand{\tr}{{\text{tr}}}

\newcommand{\iid}{\overset{\text{iid}}{\sim}}

\newcommand{\po}{{(0)}}

\makeatletter
\renewcommand\footnoterule{%
\kern-3\p@
\hrule\@width.4\columnwidth
\kern2.6\p@}
\makeatother

\makeatletter
\newcommand{\printbaselinestretch}{%
  \begingroup
    \edef\@tempa{Current \string\baselinestretch\space value: \baselinestretch}%
    \par\noindent\@tempa\par
  \endgroup
}
\makeatother

\usepackage{floatpag}

\usepackage[margin=1in]{geometry}

\usepackage{setspace}
\newif\ifsubmission

\usepackage{algorithm}
\usepackage{algorithmicx}
\usepackage{algpseudocode}

\algrenewcommand\algorithmicrequire{\textbf{Input:}}
\algrenewcommand\algorithmicensure{\textbf{Output:}}

\begin{document}

\title{\ifsubmission\else
\fi
Time-Varying Heterogeneous Treatment Effects in
Event Studies\thanks{
    \ifsubmission
\else  \texttt{botosari@mcmaster.ca} (Botosaru) and \texttt{laura.liu@pitt.edu} (Liu). We thank St\'ephane Bonhomme, Simon Freyaldenhoven, Chris Muris, Jon Roth, and conference participants at CFE-CMStatistics for helpful comments and discussions. The authors are solely responsible for any remaining errors.\fi}}
\author{Irene Botosaru \\
\textit{McMaster University} \and Laura Liu \\
\textit{University of Pittsburgh}}
\date{First version: December 11, 2024\\
This version: \today}
\maketitle
\ifsubmission\else
\fi
\begin{abstract}
    This paper examines the identification and estimation of heterogeneous treatment effects in event studies, emphasizing the importance of both lagged dependent variables and treatment effect heterogeneity. We show that omitting lagged dependent variables can induce omitted variable bias in the estimated time-varying treatment effects. We develop a novel semiparametric approach based on a short-$T$ dynamic linear panel model with correlated random coefficients, where the time-varying heterogeneous treatment effects can be modeled by a time-series process to reduce dimensionality. We construct a two-step estimator employing quasi-maximum likelihood for common parameters and empirical Bayes for the heterogeneous treatment effects. The procedure is flexible, easy to implement, and achieves ratio optimality asymptotically. Our results also provide insights into common assumptions in the event study literature, such as no anticipation, homogeneous treatment effects across treatment timing cohorts, and state dependence structure.


\end{abstract}

\newpage

\section{Introduction}

Event study methods have been a cornerstone for tracing dynamic treatment effects in empirical research across economics, finance, public policy, and related fields. Indeed, between 2020 and 2024, over thirty papers employing event study or dynamic difference-in-differences were published in the \emph{American Economic Review}. The most common implementation is via the two-way fixed-effects (TWFE) regression, which aligns units by event time rather than calendar time, allowing researchers to estimate dynamic responses to treatments and interventions, while controlling for unobserved heterogeneity that is constant over time within units (i.e., unit effects) and/or common across units within time (i.e., time effects). In practice, researchers often estimate 
\[
  Y_{it}  =  \alpha_i + \gamma_t + \sum_{j=-L}^{J} D_{it}^j \delta_j + X_{it}^\prime \beta + U_{it},
\]
where \(D_{it}^j\) indicates that unit \(i\) is \(j\) periods from its event date, \(X_{it}\) are observed covariates, \(\alpha_i\) and \(\gamma_t\) are unit and time fixed effects, and \(\{\delta_j\}\) represent average treatment effects at different leads and lags. Typically, the covariates are assumed to be strictly exogenous, i.e., they are uncorrelated with the error term across all time periods, so that current, past, and future values of the covariates do not respond to shocks in the outcome equation. This framework is attractive for its intuitive interpretation and straightforward implementation. See also recent reviews by \citet{freyaldenhoven2021visualization} and \citet{miller2023guide}.

Despite its widespread use, the standard two-way fixed effects (TWFE) estimator relies on strong assumptions that may not hold in empirical applications. In particular, by omitting lagged outcomes, it implicitly assumes that unit and time fixed effects are sufficient to eliminate all serial dependence in the residual. This assumption is often violated in settings where economic outcomes --- such as consumption, employment, earnings, and investment --- exhibit persistence due to habit formation, adjustment costs, or other dynamic mechanisms. When lagged outcomes are correlated with treatment timing, TWFE estimators conflate causal effects with residual dynamics. This can induce spurious pre-trends, bias post-treatment estimates, and lead to invalid inference, including misleading placebo tests and confidence intervals. Although dynamic panel methods are well developed, they remain underutilized in applied event study analyses.

Second, and of equal importance, is the potential heterogeneity in treatment effects. While the average treatment effect summarizes the mean response, distributional and welfare analyses often depend on the full distribution of treatment effects across units. For example, targeted subsidies may yield disproportionate benefits for specific demographic groups. Assuming homogeneous effects can mask such variation and lead to suboptimal or inequitable policy recommendations. Furthermore, treatment effects may vary systematically with observed covariates --- such as pre-treatment outcomes or demographic characteristics --- as well as unobserved unit-level attributes, including preferences or ability. Recognizing and modeling such heterogeneity is therefore essential for designing targeted interventions and for evaluating their distributional consequences.


In this paper, we introduce a semiparametric model for \emph{time-varying heterogeneous treatment effects} (TV-HTE) that simultaneously tackles outcome dynamics and cross-unit heterogeneity.  For example, we can model
\[
  Y_{it}
  = \rho_Y Y_{i,t-1} 
    + \alpha_i + \gamma_t
    + \sum_{j=0}^{J} D_{it}^j \delta_{ij}
    + X_{it}^\prime \beta
    + U_{it},\quad
  U_{it}\iid(0,\sigma_U^2),
\]
where \(\rho_Y\) captures outcome persistence,  and \(\delta_{ij}\) is the \emph{unit- and event-time-specific} treatment effect.  To reduce dimensionality, we can impose an AR($p$) process on the treatment effects. For $p=1$, we can write
\[
  \delta_{ij} = \rho_\delta \delta_{i,j-1} + \varepsilon_{ij},\quad
  \varepsilon_{ij}\iid(0,\sigma_\varepsilon^2),
  \quad j\ge1,
\]
with \(\delta_{i0}\) unrestricted.  This AR(1) specification parsimoniously captures persistence or decay in heterogeneous responses while allowing each unit to have a distinct initial effect \(\delta_{i0}\).

Interpreting \(\lambda_i=(\alpha_i,\delta_{i0})'\) as \emph{correlated random coefficients}, we permit their joint distribution to depend flexibly on the initial outcomes \(Y_{i0}\), exogenous covariates \(X_i\), and the treatment timing.  Under the assumption of conditional strict exogeneity of treatment---that $U_{it}$ is independent of treatment conditional on these covariates---and a mild non-vanishing characteristic function condition, we achieve nonparametric identification of both the common parameters \(\theta=(\rho_Y,\rho_\delta,\beta,\sigma_U^2,\sigma_\varepsilon^2)'\) and the conditional distribution of the random coefficients \(\lambda_i\).

Building on the identification result and further assuming Gaussianity on $U_{it}$ and $\varepsilon_{ij}$, we develop a two-step estimation procedure that is straightforward to implement.  In the first step, we estimate the common parameters \(\theta\) by quasi-maximum likelihood (QMLE). To do so, we assume a Gaussian form for the conditional distribution of the random coefficients \(\lambda_i\), integrate them out of the joint likelihood, and obtain \(\widehat\theta\) by maximizing the resulting marginal likelihood. We show that even when this Gaussian assumption is misspecified, the QMLE remains consistent and asymptotically normal.  

In the second step, we recover unit-specific estimates of \(\lambda_i\) via \emph
{empirical Bayes}. Let \(\widehat\lambda_i\) denote the MLE estimate of \(\lambda_i\). One can show that \(\widehat\lambda_i=\lambda_i+V_i\), where $V_i$ 
has mean zero and a variance matrix estimated from the first-step output. 
Tweedie's formula then yields the posterior mean that combines this noisy MLE estimate with a correction term that depends on the derivative of the marginal density of the sufficient statistics. Intuitively, this correction shrinks the MLE estimate toward regions of higher density in the data, effectively combining information across units to improve the estimation accuracy. 

By focusing on the derivative of the observed marginal density of the sufficient statistics \(p(\widehat\lambda_i\mid Y_{i0},X_i)\), we sidestep the challenging deconvolution problem to recover the underlying distribution of \(\pi(\lambda\mid Y_0,X)\). The marginal density of the sufficient statistics can be estimated either parametrically or nonparametrically, and the resulting empirical Bayes estimator shrinks noisy unit-level estimates toward a data-driven prior and achieves ratio optimality, that is, its compound risk converges to the oracle risk that would be attained by an infeasible estimator with perfect knowledge of the true conditional random coefficient distribution.

This TV-HTE framework provides several advantages compared to the standard event study methods.  Incorporating the lagged dependent variable eliminates omitted-variable bias due to persistence.  Modeling heterogeneity through a time-series process captures the dynamics in treatment effects without high-dimensional estimation.  The empirical Bayes step sharpens unit-level estimates in short panels, overcoming the many-means problem.  

In addition to the above setup, our framework extends naturally to discrete or continuous treatments and to staggered adoption designs.  We also allow for both strictly exogenous covariates, whose coefficients may be unit-specific or common, and predetermined covariates with common effects.  The dynamics for $Y_{it}$ and $\delta_{ij}$ can be generalized to AR($p$) processes, e.g., AR(2) to capture oscillatory patterns, and the error structure can be generalized to allow for cross-sectional heteroskedasticity $\sigma_{U,i}^2$ or MA($q$) process.

Moreover, our framework also sheds light on common assumptions in event study.  For example, by examining the estimated means of the event-time coefficients in pre-treatment periods ($j<0$), we can formally test the no anticipation assumption.  Also, by comparing these means across cohorts defined by treatment timing, we can assess the homogeneity of treatment effects. In addition, our dynamic panel structure with separate persistence parameters for the outcome $\rho_Y$ and the treatment effects $\rho_\delta$ allows us to evaluate state dependence in both the underlying process and the policy response.

We assess the performance of our TV-HTE estimator through extensive Monte Carlo experiments and an empirical example on county-level unemployment during the 2008 Great Recession.  In the Monte Carlo, our method nearly replicates the infeasible oracle in recovering the distribution of unit-specific effects under Gaussian, bimodal, and heavy-tailed distributions, and across dynamic response profiles ranging from monotonic decay to oscillatory paths. Our tests maintain correct size under the null and exhibit high power.  In the empirical example, we find that the heterogeneous treatment effects are markedly non-Gaussian and irregularly distributed: county-level unemployment spikes range from roughly 0.5 to over 7 percentage points, far surpassing the average TWFE estimate, and dynamic trajectories differ across counties as well. Formal tests reject the random effects specification, the null of no correlation between heterogeneous effects and baseline heterogeneity, and the null of no state dependence, instead supporting our correlated random coefficients, time-varying analysis.

\paragraph{Related literature.} 

Since the pioneering work by \citet{ashenfelter1978estimating} on estimating the effects of training programs on earnings using a two-way fixed-effects model, empirical researchers have widely adopted panel data event study designs to quantify causal effects in economics. However, a growing literature recognizes that homogeneous effect assumptions can yield misleading estimates in staggered adoption settings, and recent work has fallen into three methodological strands. First, robust estimators for the mean treatment effect, such as \citet{dechaisemartin2023twoway} and \citet{borusyak2024revisiting}, rely on carefully constructed two-by-two comparisons or imputation-based counterfactuals to eliminate bias. Second, group-level approaches, such as \citet{callaway2021difference}, \citet{goodman2021difference}, and \citet{dechaisemartin2023twoway}, estimate cohort- and period-specific treatment effects and aggregate them with convex weights or interaction weighted regressions to ensure no negative contributions. Finally, \citet{arkhangelsky2024design} consider individual-level treatment effects via finite-mixture and latent-type models. In this paper, we also examine individual-level treatment effects and incorporate an empirical Bayes approach to refine these estimates, thereby improving precision while flexibly accommodating time-varying heterogeneity. Our analysis also helps assess common assumptions underlying event study designs, such as those in \citet{sun2021estimating}.

To accommodate outcome persistence and mitigate the Nickell bias in short panels, we draw on dynamic panel methods. \citet{anderson1982formulation} propose first-differencing and using deeper lags as instruments to eliminate fixed effects. \citet{arellano1991some} generalize this with a GMM estimator that exploits all available lagged levels, substantially improving efficiency in panels with small $T$. \citet{blundell1998initial}’s system GMM further addresses weak-instrument concerns when the autoregressive coefficient is high. \citet{arellano2012identifying} show that, under mild serial-correlation restrictions, one can identify moments—and even the full distribution—of random coefficients in a short panel. \citet{alvarez2022robust} develop robust QMLE for dynamic panels that remain valid under heteroskedasticity and arbitrary serial correlation, demonstrating that random-effects likelihood methods can outperform GMM when distributional assumptions approximately hold. In this paper, we similarly estimate the common autoregressive parameters via QMLE in the first step, and the time dynamics of the heterogeneous treatment effects are further modeled by time-series processes to reduce dimensionality. 

Our second step employs an empirical Bayes estimator to recover unit-specific treatment trajectories. \citet{robbins1951asymptotically} introduces empirical Bayes as a compound decision problem, yielding shrinkage rules that minimize average risk without knowing the prior distribution. With exponential family likelihood, Tweedie's formula links posterior means to the derivatives of the marginal density of sufficient statistics, enabling nonparametric $\pi$-modeling empirical Bayes \citep{efron2011tweedie}. \citet{brown2009nonparametric} and \citet{jiang2009general} establish that maximum-likelihood empirical Bayes estimators for normal-means problems achieve asymptotic minimaxity or ratio optimality. \citet{gu2017unobserved} and \citet{liu2020forecasting} show substantial gains in estimation and forecasting accuracy by efficiently combining information across cross-sectional units. In this paper, we employ both parametric and nonparametric empirical Bayes to obtain posterior mean estimates of unit-specific treatment trajectories, optimally balancing individual signal and noise, and establish their ratio optimality.

The remainder of this paper is organized as follows. Section \ref{sec:model} introduces the model and discusses the identification of time-varying heterogeneous treatment effects. Section \ref{sec:estimation} presents our two-step estimation method and establishes its asymptotic properties, including ratio optimality.
Section \ref{sec:extensions-test} extends our estimator to various contexts and discusses tests for common event study assumptions. 
Section \ref{sec:sim} conducts Monte Carlo experiments to examine the finite-sample properties of our estimators. Section \ref{sec:app} employs our panel data estimator to analyze how the Great Recession in 2008 affected local labor markets. Finally, Section \ref{sec:conclusion} concludes. Appendix \ref{sec:proofs} provides the proofs for all propositions and theorems, and the online appendix contains additional tables and figures. 

\section{Simple model and identification}\label{sec:model}
\subsection{Importance of lagged dependent variables}\label{sec:lagged-dep}

Economic series tend to be persistent over time.  For example, consumption adjusts gradually as habits evolve, and wages move slowly amid contract and adjustment frictions.  When such built-in persistence coincides with event timing, the dummy variables in a TWFE regression absorb not only the true effect of the intervention but also the persistence present in the data.  As a result, what appear as treatment effects may also reflect the persistence of past outcomes, giving rise to spurious pre-trends, distorted post-treatment estimates, and misleading inference in placebo tests and confidence intervals.

A simple, yet revealing, illustration shows why excluding lagged dependent variables from an event study regression generates omitted variable bias in the estimated treatment effect path.  Consider a panel with five periods ($t=0,1,2,3,4$) and a common treatment occurring at $t=2$, so that $D_{it}^j=\mathbf1\{t-j=2\}$.  Suppose the true DGP is an AR(1) model with persistence $\rho_Y$ and a treatment effect path $(\delta_0,\delta_1,\delta_2)$,
\[
  Y_{it}  =  \rho_Y Y_{i,t-1}
     + \sum_{j=0}^2 D_{it}^j \delta_j
     + U_{it},
\]
and let $\E[Y_{i0}]=0$ for simplicity. 
In contrast, the naive event study regression omits dynamics and simply fits
\[
  Y_{it}
  = \sum_{j=0}^2 D_{it}^j \widetilde\delta_j
    + \widetilde U_{it}.
\] 
Because the true outcomes are serially correlated, each indicator $D_{it}^j$ is correlated with the omitted lag $Y_{i,t-1}$, producing bias in $\widetilde\delta_j$.  One can show analytically that
\[
  \mathrm{Bias}(\widetilde\delta_j)
   = \rho_Y \E\left[D_{it}^j Y_{i,t-1}\right]
   = \rho_Y \E[Y_{i,j+1}]
   = 
    \begin{cases}
      0, & j=0,\\
      \rho_Y \delta_0, & j=1,\\
      \rho_Y \delta_1  + \rho_Y^2 \delta_0, & j=2.
    \end{cases}
\]
Thus, even if the true effect at $j=0$ is identified without bias, biases accumulate at longer horizons, distorting the entire treatment path.  

\begin{figure}[t]
  \caption{Omitted variable bias - toy example}
  \label{fig:ovb}
  \begin{center}
      \vspace{-1em}
  \includegraphics[width=.7\textwidth]{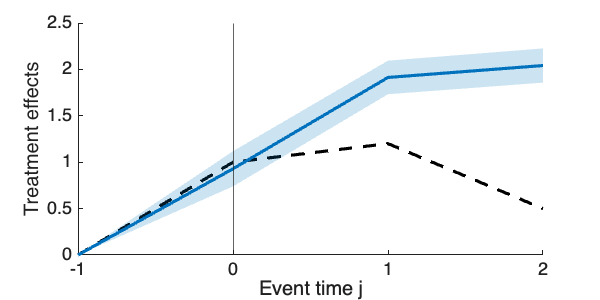}
      \vspace{-1em}
  \end{center}
  {\footnotesize {\em Notes:} The black dashed line shows the true treatment effect path $(\delta_0,\delta_1,\delta_2)=(1,1.2,0.5)$, while the blue solid line shows the estimated treatment effect path of $\{\widetilde\delta_j\}$ from a naive event study regression without lagged dependent variables. The blue band shows the 95\% confidence interval.
  }\setlength{\baselineskip}{4mm}
  \end{figure}

Figure \ref{fig:ovb} contrasts the true effects (black dashed) with the biased estimates (blue solid) for $\rho_Y=0.8$ and $(\delta_0,\delta_1,\delta_2)=(1,1.2,0.5)$ in a simulated sample of $N=100$, and their differences are statistically significant.  This toy example highlights the necessity of explicitly modeling lagged dynamics in event study designs.  By incorporating $Y_{i,t-1}$, researchers can control for outcome persistence and recover unbiased estimates of the time-varying treatment effects.  

\subsection{Dynamic panel with time-varying het.\ treatment effects}\label{sec:model-dynamic}

We now introduce a simple dynamic panel framework that accommodates both persistence in the outcome and heterogeneous treatment effects across units and event time horizons. To highlight the main intuition, we focus on a simple model that drops time fixed effects and other covariates, and adopts a common treatment timing in this section. More general cases are discussed in subsequent sections.

Let \(i = 1,\dots,N\) index cross-sectional units and \(t = 0,\dots,T\) denote time periods.  We consider a large \(N\), fixed \(T\) setup, which is natural for many event study applications where the number of treated and control units is large but the available pre- and post-treatment windows are of limited length. For simplicity, each unit undergoes a single treatment at a common period \(t_0\).  We define the event time indicator
\(
  D_{it}^j  =  \mathbf1\{ t - j = t_0\},\quad
  j = 0,1,\dots,J,
\)
so that \(D_{it}^j=1\) when unit \(i\) is in the \(j\)th period after treatment.  Our baseline outcome equation augments a standard dynamic panel with these event time dummies
\begin{equation}\label{eq:dyn_panel}
  Y_{it}
  = \rho_Y Y_{i,t-1}
    + \alpha_i
    + \sum_{j=0}^{J} D_{it}^j \delta_{ij}
    + U_{it},
    \quad
  U_{it} \iid (0,\sigma_U^2).
\end{equation}
Here, \(\rho_Y\) captures first-order persistence in the outcome, while the unit-specific intercept \(\alpha_i\) controls for time-invariant heterogeneity.  The term \(\delta_{ij}\) is the treatment effect for unit \(i\) at event time \(j\), allowing each unit to respond differently and dynamically to the intervention.

Because freely estimating the full matrix \(\{\delta_{ij}\}\) would involve \((J+1)\times N\) parameters, we can incorporate a simple time series structure on the heterogeneous effects to reduce the dimensionality.\footnote{The assumed time series structure for $\delta_{ij}$ is testable in the data. For example, one can obtain preliminary estimates of the individual effect trajectories by orthogonal forward differencing of \citet{arellano1995another}, and then subject these series to standard time-series diagnostics to assess whether an AR($p$) process provides an adequate fit.}  For example, for \(j\ge1\) we assume an AR(1) process
\begin{equation}\label{eq:delta_ar1}
  \delta_{ij}
  = \rho_\delta \delta_{i,j-1}
    + \varepsilon_{ij},
  \quad
  \varepsilon_{ij} \iid (0,\sigma_\varepsilon^2).
\end{equation}
The persistence parameter \(\rho_\delta\) governs the decay or oscillation of treatment effects over successive periods, while the variance \(\sigma_\varepsilon^2\) captures unit-specific shocks to the response path.  Only the initial effect \(\delta_{i0}\) remains freely heterogeneous, enabling each unit to have its own starting point for the dynamic treatment response.

To capture potential correlations between initial outcomes, individual heterogeneity, and initial treatment effects, we let
\[
  \lambda_i  =  (\alpha_i,\delta_{i0})',
  \quad
  \lambda_i \mid Y_{i0} \sim  \pi(\lambda_i\mid Y_{i0}),
\]
where \(\pi(\lambda\mid Y_0)\) is an unrestricted conditional density.  This correlated random coefficients specification allows \(\alpha_i\) and \(\delta_{i0}\) to depend flexibly on the initial outcome \(Y_{i0}\) (and, in extensions, on additional exogenous covariates). Moreover, by allowing for correlation between the baseline heterogeneity \(\alpha_i\) and the initial treatment effects \(\delta_{i0}\), the framework can capture meaningful heterogeneity in treatment effects that standard event study methods might overlook.

Collecting the parameters into the vector
\(
  \theta = (\rho_Y,\rho_\delta,\sigma_U^2,\sigma_\varepsilon^2)'
\)  with true value $\theta_0$,
we aim to recover \(\theta\), the conditional distribution of \(\lambda_i\), and posterior mean estimates of  \(\lambda_i\).

\subsection{Identification}\label{sec:id}

We now formalize the conditions under which both the common parameters 
\(\theta\) 
and the conditional distribution of the unit-specific coefficients 
\(\lambda_i \) are nonparametrically identified.  

\begin{assumption}[Model]\label{assu:model} Consider the simple model given by \eqref{eq:dyn_panel} and \eqref{eq:delta_ar1} with common treatment period $t_0$. 
  \begin{enumerate}[label=(\alph*)]
        \item  \((Y_{i0},\lambda_i)\) are i.i.d.\ across \(i\).  
        \item 
        $U_{it}\perp(Y_{i,0:t-1},\lambda_i)$, $\varepsilon_{ij}\perp(\delta_{i,1:j-1},Y_{i0},\lambda_i)$, and $U_{it}\perp\varepsilon_{ij}$, for all $i$, $t$, and $j$.
      \end{enumerate}
  \end{assumption}
Condition (b) implies that the combined error terms \(\check U_{i,1:T}(\rho_\delta)\) in \eqref{eq:Y_it} and hence the noise \(V_i(\rho_\delta)\) in \eqref{eq:suff-stat} below are independent of \(\lambda_i\) conditional on \(Y_{i0}\), a key requirement for the deconvolution exercise. 

\begin{remark}[Conditional exogeneity in treatment]\label{rmk:cond-exo}
\normalfont{
Under a common treatment timing, our baseline specification implicitly imposes conditional exogeneity of treatment: the innovation $U_{it}$ is assumed independent of the event time indicators $D_{it}^j$ (or, in a more general case with different treatment timings, independent once we condition on observed covariates).  This condition ensures that the design matrix $W_i(\rho_\delta)$ for heterogeneous coefficients in \eqref{eq:W_i} below is exogenous, so that the deconvolution step yields valid identification results.

It is useful to contrast this with the classic parallel trends assumption, which typically requires no outcome persistence ($\rho_Y=0$) and $\E[U_{it}^\po\mid\{D_{ij}^j\}]=0$, where $U_{it}^\po$ denotes the potential error under no treatment. Here we relax the parallel trends assumption by allowing $\rho_Y\neq0$.\footnote{Under our model, the transformed outcome $Y_{it}-\rho_Y Y_{i,t-1}$ satisfies a conditional parallel trend assumption once we control for exogenous covariates, as discussed in \citet{wooldridge_two-way_2021}.} 
Although our conditional exogeneity assumption is stronger than standard parallel trends in terms of its assumption on the error terms, it affords us the flexibility to estimate richer heterogeneous treatment effect trajectories.  

Moreover, by framing $(\alpha_i,\delta_{i0})$ as correlated random coefficients, we naturally accommodate \emph{selection on unobservables}, where treatment timing can correlate with observed covariates, latent heterogeneity including heterogeneous treatment effects, as well as time fixed effects in the general model.
}
\end{remark}

Combining the simple dynamic panel data model \eqref{eq:dyn_panel} and the AR(1) process \eqref{eq:delta_ar1}, we obtain
\begin{align}
  Y_{it} - \rho_Y Y_{i,t-1}
  = \alpha_i
    + \left(\sum_{j=0}^{J} \rho_\delta^j D_{it}^j\right) \delta_{i0}
    + \underbrace{\left(U_{it}
              + \sum_{j=0}^{J}\sum_{k=1}^j \rho_\delta^{ j-k} D_{it}^j \varepsilon_{ik}\right)}
      _{\equiv\check U_{it}(\rho_\delta)}.\label{eq:Y_it}
\end{align}
where \(\check U_{i,1:T}(\rho_\delta)\) is a mean-zero vector with covariance matrix \(\Sigma_{\check U}(\theta)\). Next, define the \(T\times2\) design matrix \(W_i(\rho_\delta)\) by
  \begin{align}
    W_i(\rho_\delta)
    = \begin{pmatrix}
        1 & \sum_{j=0}^{J} \rho_\delta^j D_{i1}^j \\
        1 & \sum_{j=0}^{J} \rho_\delta^j D_{i2}^j \\
        \vdots & \vdots \\
        1 & \sum_{j=0}^{J} \rho_\delta^j D_{iT}^j
      \end{pmatrix},\label{eq:W_i}
  \end{align}
and let \(W_{it}(\rho_\delta)\) be its \(t\)-th row.\footnote{In our simple setup with common treatment timing $t_0$, the design matrix $W_i(\rho_\delta)$ is deterministic and homogeneous across all units, so there is no need to condition on it in the assumptions and derivations, thereby simplifying the exposition.\label{fn:no-W}} The model can then be written compactly as
\[
  Y_{it} - \rho_Y Y_{i,t-1}
  = W_{it}(\rho_\delta)' \lambda_i
    + \check U_{it}(\rho_\delta).
\]
Given \(\rho=(\rho_\delta,\rho_Y)'\), the OLS/MLE estimator of the latent coefficient vector \(\lambda_i\) is
\begin{align}
  \widehat\lambda_i(\rho)
  = W_i(\rho_\delta)^+ \left(Y_{i,1:T} - \rho_Y Y_{i,0:T-1}\right) = \lambda_i + V_i(\rho_\delta),\label{eq:suff-stat}
\end{align}
where $W_i(\rho_\delta)^+ 
= \left(W_i(\rho_\delta)'W_i(\rho_\delta)\right)^{-1}
  W_i(\rho_\delta)'$ and $V_i(\rho_\delta)
  = W_i(\rho_\delta)^+\check U_{i,1:T}(\rho_\delta),$ which is mean-zero and has covariance matrix $\Sigma_{V,i}(\theta) = W_i(\rho_\delta)^+ \Sigma_{\check U}(\theta) [W_i(\rho_\delta)^+]'$.  
Thus, \(\widehat\lambda_i(\rho)\) is a sufficient statistic for \(\lambda_i\) with noise \(V_i(\rho_\delta)\).

\begin{assumption}[Distributions]\label{assu:dist}
\leavevmode
\begin{enumerate}[label=(\alph*)]
      \item The characteristic functions of \(\lambda_i\mid Y_{i0}\), \(U_{it}\), and \(\varepsilon_{ij}\) are non-vanishing almost everywhere. 
      \item The characteristic functions of \(U_{it}\) and \(\varepsilon_{ij}\) are twice differentiable.
      \item $\Var(\delta_{i0}) > 0$ and $\Var(Y_{i0}) > 0$.
    \end{enumerate}
\end{assumption}
Conditions (a) and (b) guarantee that the convolution in \eqref{eq:suff-stat} can be inverted via characteristic function methods, thereby recovering the conditional distribution of \(\lambda_i\mid Y_{i0}\).  Condition (c) ensures cross-sectional variation in both the initial treatment effects and initial outcomes, guaranteeing that the moment conditions for identifying $\rho_\delta$ and $\rho_Y$ are non-degenerate.

\begin{assumption}[Rank condition]\label{assu:rank} 
      $t_0 \ge3$, and $T - t_0 \ge J \ge 1$.
  \end{assumption}
Since \(\check U_{i,1:T}(\rho_\delta)\) is an MA(\(J\)) process in the error terms
\(\{U_{it},\varepsilon_{ij}\}\), we require sufficient pre-treatment variation to disentangle these shocks from the treatment effect dynamics. In the simple common timing design, this amounts to imposing $t_0 \ge3$, which helps satisfy the rank conditions in \citet{arellano2012identifying}. For general cases with different treatment timings and additional covariates, we can extend to a more general rank condition on the expanded design matrix. $T - t_0 \ge J \ge 1$ ensure that there are enough post-treatment observations to identify the full sequence of dynamic treatment effects.

\begin{theorem}[Nonparametric identification]\label{thm:ident}
Under Assumptions \ref{assu:model}--\ref{assu:rank}, the common parameters \(\theta\) and the conditional density \(\pi(\lambda_i\mid Y_{i0})\) are identified.
\end{theorem}
First, we can identify the autoregressive parameters $\rho$ from moment conditions. Second, the identification of the conditional density \(\pi(\lambda_i\mid Y_{i0})\) relies on the sufficient statistics representation \eqref{eq:suff-stat}.  Taking characteristic functions on both sides transforms the convolution in the time domain into a product in the frequency domain, so one obtains on the right hand side a product of the characteristic functions of the latent coefficients $\lambda_i\mid Y_{i0}$ and the noise term $V_i(\rho)$.  Under the non-vanishing characteristic functions, this product can be deconvolved to recover both distributions. Our proof builds on the deconvolution argument of \citet{arellano2012identifying} and \citet{liu2023density} for correlated random coefficients panels and extends it to the dynamic event study framework.

\section{Estimation and asymptotics}\label{sec:estimation}
\subsection{Two-step estimation}\label{sec:estimation-2step}

Building on the identification results and further assuming Gaussianity on $U_{it}$ and $\varepsilon_{ij}$, we implement a simple two-step estimator that first estimates the common parameter and then recovers the unit-specific parameters, as summarized in Algorithm \ref{alg:TV-HTE}.  

\begin{algorithm}[t]
  \begin{onehalfspace}
\caption{Semiparametric TV-HTE estimator}\label{alg:TV-HTE}
\begin{algorithmic}
\Require Panel data $\{Y_{it}\}_{i=1,\dots,N}^{t=0,\dots,T}$, treatment timing $t_0$, horizon $J$.
\Ensure Estimates of common parameters $\widehat\theta$ and unit-level parameters $\{\widetilde\lambda_i\}$.

\medskip
\State \textbf{Step 1: QMLE for common parameters.} Maximize the marginal quasi-log-likelihood
\[
\ell_N(\theta,b_0,b_1,\Sigma_\lambda)
=\sumi
\log\phi\left(Y_{i,1:T};\,\mu_i(\theta,b_0,b_1),\,\Omega_i(\theta,\Sigma_\lambda)\right),
\]
where $\mu_i(\cdot)$ and $\Omega_i(\cdot)$ are given in \eqref{eq:mlik-mu-omega},
to obtain
    \(\left(\widehat\theta,\widehat b_0,\widehat b_1,\widehat\Sigma_\lambda\right)\). 

\medskip
\State \textbf{Step 2: Empirical Bayes for unit-specific parameters}
\begin{enumerate}
  \item Build $T\times2$ matrix \(\widehat W_i=W_i(\widehat\rho_{\delta})\) with rows
    \(\left[1,\sum_{j=0}^{J}\widehat\rho_\delta^j D_{it}^j\right]\), and $\widehat W_i^+=\left(\widehat W_i'\widehat W_i\right)^{-1}\widehat W_i'$.
  \item Compute OLS/MLE estimate and noise covariance
    \[
      \widehat\lambda_i
      = \widehat W_i^+\left(y_{i,1:T} - \widehat\rho_Y y_{i,0:T-1}\right),
      \quad
      \widehat\Sigma_{V,i} = \widehat W_i^+ \Sigma_{\check U}(\widehat\theta) \widehat W_i^{+\prime}.
    \]
  \item Estimate marginal density of the sufficient statistics
    \(p(\widehat\lambda_i\mid Y_{i0})\) 
    either parametrically or nonparametrically.
  \item Apply Tweedie's formula:
    \[
      \widetilde\lambda_i
      = \widehat\lambda_i
        + \widehat\Sigma_{V,i}         \nabla_{\widehat{\lambda}_i}\log \widehat p
        \left(\widehat\lambda_i\mid y_{i0}\right).
    \]
\end{enumerate}
\end{algorithmic}
\end{onehalfspace}
\end{algorithm}

In the first step, we estimate the common parameters $\theta$ by QMLE, treating the latent coefficients $\lambda_i\mid Y_{i0}$ as if they followed a Gaussian regression model
\[
  \lambda_i\mid Y_{i0} \sim 
  N\left(b_0 + b_1 Y_{i0},\,\Sigma_\lambda\right).
\]
Even though this correlated random coefficients distribution may be misspecified, maximizing the resulting marginal likelihood over $\theta$ and the nuisance parameters $(b_0,b_1,\Sigma_\lambda)$ yields consistent and asymptotically normal estimates for $\theta$. In practice, the Gaussian prior and likelihood imply conjugacy, yielding a closed-form marginal likelihood 
\[
  Y_{i,1:T} \sim 
  \mathcal N\left(
    \mu_i(\theta,b_0,b_1),
     \Omega_i(\theta,\Sigma_\lambda)
  \right),
\]
where
\begin{align}
  \mu_i(\theta,b_0,b_1)
  &= A(\rho_Y) Y_{i0} + \widetilde{W}(\rho_Y,\rho_\delta) \left(b_0 + b_1Y_{i0}\right), \label{eq:mlik-mu-omega}\\
  \Omega_i(\theta,\Sigma_\lambda) 
  &= B(\rho_Y) \Sigma_{\check U}(\theta) B(\rho_Y)' + \widetilde{W}(\rho_Y,\rho_\delta) \Sigma_\lambda \widetilde{W}(\rho_Y,\rho_\delta)', \notag
\end{align} 
where $A(\rho_Y) = (\rho_Y, \rho_Y^2, \ldots, \rho_Y^{T})'$ captures initial condition propagation, $B(\rho_Y)$ is the $T \times T$ lower triangular matrix with $(s,t)$-th element $\rho_Y^{s-t}$ for $s \geq t$ (zero otherwise), and $\widetilde{W}(\rho_Y,\rho_\delta) = B(\rho_Y)W(\rho_\delta)$ transforms the treatment design matrix. We can efficiently maximize this marginal likelihood using standard numerical optimization routines.

In the second step, the sufficient statistic $\widehat\lambda_i(\rho)$ has been derived in Section \ref{sec:id}: see equation \eqref{eq:suff-stat}.  For the empirical Bayes estimator, we exploit Tweedie's formula \citep{robbins1951asymptotically,efron2011tweedie} to compute the posterior mean of each unit's random coefficients \(\lambda_i\), 
\begin{align}
  \E\left[\lambda_i\mid Y_{i,0:T},t_0,\rho,p\right]
  = \widehat\lambda_i(\rho)
    + \Sigma_{V,i}(\theta)       \frac{\partial}{\partial\widehat\lambda_i(\rho)}
      \log p\left(\widehat\lambda_i(\rho)\mid Y_{i0}\right).\label{eq:tweedie}
\end{align}
The first term is the OLS/MLE estimate and the sufficient statistic $\widehat\lambda_i(\rho)$, while the second term is a Bayes correction that depends on the derivative of the marginal density of the sufficient statistics \(\widehat\lambda_i(\rho) \mid Y_{i0}\).  
The correction term adapts to the local shape of the marginal density of \(\widehat\lambda_i(\rho) \mid Y_{i0}\): a positive derivative indicates the estimate falls below the mode so we shrink upward, while a negative derivative indicates it lies above the mode so we shrink downward. Moreover, steeper slopes, i.e., higher density concentration, yield larger corrections, whereas flatter regions induce milder shrinkage.

With fixed \(T\) in event studies, the unit-specific parameters $\lambda_i$ cannot be consistently estimated; instead, the empirical Bayes estimator helps efficiently combine information across all units to shrink and refine these estimates, thereby reducing the overall compound risk.  Crucially, Tweedie's formula circumvents the challenge to deconvolve the latent coefficient density \(\pi(\lambda_i\mid Y_{i0})\); one only needs to estimate the marginal density of the observable quantities \(\left(\widehat\lambda_i(\rho),Y_{i0}\right)\).\footnote{Since the conditional and joint log densities differ only by a constant that drops out under differentiation, we can work with $\log p\left(\widehat\lambda_i,Y_{i0}\right)$ instead of $\log p\left(\widehat\lambda_i\mid Y_{i0}\right)$ in practice.}  In practice, this marginal can be fit parametrically, such as plugging in the Gaussian form implied by the QMLE, or nonparametrically via kernel or mixture methods. The former is easier to implement, while the latter helps reveal richer heterogeneity patterns.  The resulting empirical Bayes estimator shrinks the noisy OLS/MLE $\widehat\lambda_i(\rho)$ toward a data-driven prior and attains ratio optimality, i.e., its compound risk is asymptotically equivalent to the oracle risk, where one knows the true conditional distribution of \(\lambda_i\).

\subsection{Asymptotics for QMLE}\label{sec:QMLE}

We now establish that the QMLE in the first step is consistent and asymptotically normal.
\begin{assumption} (Estimation)\label{assu:est}
  \begin{enumerate}[label=(\alph*)]
  \item \(U_{it}\) and \(\varepsilon_{ij}\) follow Gaussian distributions with $\sigma_U^2,\sigma_{\varepsilon}^2>0$.
  \item $(\lambda_{i},Y_{i0})$ have finite fourth moment.
  \end{enumerate}
\end{assumption}
This Gaussianity condition (a) is imposed for the two-step estimator, not for identification.  Nonparametric identification in Theorem \ref{thm:ident} only requires a non-vanishing characteristic function of the composite noise, regardless of its exact distribution.  In more general specifications with additional covariates, we need only conditional Gaussianity of \(\{U_{it},\varepsilon_{ij}\}\) given those covariates.  Furthermore, if one forgoes the AR(\(p\)) dimension reduction and instead directly estimates the full vector of \(\{\delta_{ij}\}\), the normality of \(\varepsilon_{ij}\) can also be dispensed with.  However, when employing the AR-based reduction, where \(V_i(\rho)\) is a linear combination of \(U_{it}\) and \(\varepsilon_{ij}\), we require that this composite noise lie in an exponential family, such as Gaussian, to obtain the Tweedie's formula for the empirical Bayes estimator. 

Let \(\eta=(\theta',\,b_0',\,b_1',\,\mathrm{vech}(\Sigma_\lambda)')'\) collect both the common parameters and the Gaussian prior parameters, and $\eta_0$ be the pseudo-true value of $\eta$.
For the prior parameters, $b_{0,0}$ and $b_{1,0}$ are those that minimize the Kullback-Leibler distance between the true conditional distribution of $\lambda_i\mid Y_{i0}$ and the working Gaussian regression. Equivalently, $b_{1,0}$ is the best linear predictor coefficient of $\lambda_i$ on $Y_{i0}$ and $b_{0,0}=\E[\lambda_i]-b_{1,0}\E[Y_{i0}]$, while $\Sigma_{\lambda,0}$ is the corresponding residual covariance. 

\begin{theorem}[QMLE]\label{thm:QMLE}
Under Assumptions \ref{assu:model}-\ref{assu:rank} and \ref{assu:est},  
\[
  \widehat\eta
   \xrightarrow{p} \eta_0,
  \quad
  \sqrt{N}\left(\widehat\eta-\eta_0\right)
   \xrightarrow{d} 
  \mathcal{N}\left(0,\,
    H(\eta_0)^{-1} G(\eta_0)  H(\eta_0)^{-1}
  \right),
\]
where 
\[
  H(\eta_0)
  = -\E\left[\nabla^2_{\eta}\ell_i(\eta_0)\right],
  \quad
  G(\eta_0)
  = \E\left[\nabla_{\eta}\ell_i(\eta_0) \nabla_{\eta}\ell_i(\eta_0)'\right],
\]
and \(\ell_i\) is the marginal quasi-log-likelihood of \(Y_{i,1:T}\).  The asymptotic variance of \(\widehat\theta\) is obtained by taking the corresponding sub-block of this sandwich matrix.  
\end{theorem}
The intuition is in line with standard M-estimation arguments applied to a pseudo-likelihood: the identification and moment conditions ensure a unique maximizer and uniform convergence of the score, while smoothness guarantees a valid Taylor expansion of the log-likelihood.  The resulting sandwich-form variance reflects potential misspecification of the prior.  Note that the there is no Nickell bias for the marginal likelihood after integrating out $\lambda_i$, although there is for the conditional likelihood: see also the robust QMLE discussion in \citet{alvarez2022robust}.
    
\subsection{Ratio optimality for empirical Bayes}\label{sec:empirical-bayes}
In this subsection, we show that the empirical Bayes estimator in the second step achieves oracle risk performance.

Define the risk for any estimator $\widetilde\lambda_{1:N}$ and the oracle risk as follows:
\[
  R_N(\widetilde\lambda_{1:N};\,\theta_0,\pi_0)=\E_{\theta_0,\pi_0}\left[\sum_{i=1}^N\|\widetilde\lambda_i-\lambda_i\|^2\right],
  \quad
  R_N^{\mathrm{oracle}}(\theta_0,\pi_0)=\E_{\theta_0,\pi_0}\left[\sum_{i=1}^N \Var_{\theta_0,\pi_0}(\lambda_i\mid Y_{i,0:T})\right],
\]
where the subscripts $(\theta_0,\pi_0)$ indicate that the expectation and variance are under the true data generating law $\P_{\theta_0,\pi_0}$. $\theta_0$ and $\pi_0$ are unknown to the econometrician but fixed in the DGP. Let the leave-one-out kernel estimator be
\[
\widehat p_{(-i)}(\widehat\lambda_i(\rho),y_{i0})=
\frac{1}{N-1}\sum_{j\ne i}\frac{1}{B_N^d}
\phi\left(\tfrac{\widehat\lambda_j(\rho)-\widehat\lambda_i(\rho)}{B_N}\right)
\phi\left(\tfrac{Y_{j0}-y_{i0}}{B_N}\right),
\]
with $d=\dim(\widehat\lambda)+1$, and the empirical Bayes estimator for $\lambda_i$ be
\begin{align}
  \widetilde \lambda_i=\left[\widehat\lambda_i(\widehat\rho)+\left(\widehat\Sigma_{V,i}+B_N^2 I_{\dim\left(\widehat\lambda_i\right)}\right)\frac{\partial}{\partial\widehat\lambda_i(\widehat\rho)}
  \log \widehat p_{(-i)}\left(\widehat\lambda_i(\widehat\rho)\mid Y_{i0}\right)\right]_{C_N},\label{eq:lambda-tilde}
\end{align}
where $\widehat\Sigma_{V,i}$ is given in Algorithm \ref{alg:TV-HTE}, and $[\cdot]_{C_N}$ means truncate the vector inside to lie within the Euclidean ball of radius $C_N$.  

We adopt Assumptions 3.2--3.6 of \citet{liu2020forecasting}, restated as Assumptions \ref{assu:trimbandwidth}--\ref{assu:thetahat} in Appendix \ref{sec:proof-ratio-optimality}. 
First, exponential tails for $(\lambda_i,Y_{i0})$ ensure that the probability mass trimmed away at $\|\lambda_i\|>C_N$ vanishes as $N\to\infty$.
Second, trimming and bandwidth rates $(C_N,C'_N,B_N)$ balance kernel bias and variance.
Third, smoothness of $\pi(Y_{i0}\mid\lambda_i)$ prevents sharp spikes in the distribution of $Y_{i0}$.  Together, these conditions ensure that the leave-one-out density $\widehat p_{(-i)}$ is consistent.
Fourth, posterior-mean truncation ensures that the empirical Bayes procedure remains stable by preventing outlier units with extreme estimates from dominating the overall performance, thereby maintaining uniform control over the risk across all possible priors.
Finally, $\sqrt N$-consistency of the common parameters $\widehat\theta$ follows from the QMLE result in Theorem \ref{thm:QMLE}.

\begin{theorem}[Ratio optimality]\label{thm:ratio-optimality}
  Let $\theta_0$ denote the unknown true parameter, treated as fixed in the DGP. Under Assumptions \ref{assu:model}--\ref{assu:rank}, \ref{assu:est}, and \ref{assu:trimbandwidth}--\ref{assu:thetahat}, the empirical Bayes estimator $\widetilde\lambda_{1:N}$ in \eqref{eq:lambda-tilde} achieves $\varepsilon_0$-ratio optimality uniformly over $\pi_0\in\Pi$: for any $\varepsilon_0>0$,
  \[
    \limsup_{N\to\infty}\sup_{\pi_0\in\Pi}
    \frac{R_N(\widetilde\lambda_{1:N};\,\theta_0,\pi_0)-R_N^{\mathrm{oracle}}(\theta_0,\pi_0)}{N\E_{\theta_0,\pi_0}[\Var_{\theta_0,\pi_0}(\lambda_i\mid Y_{i,0:T})]+N^{\varepsilon_0}}
    \le0.
  \]
\end{theorem}
In a decision theoretic framework for compound risk, our event study estimator attains ratio optimality, meaning that its overall risk converges to the infeasible oracle benchmark up to vanishing terms.  In other words, the mean squared error of our empirical Bayes shrinkage estimator is asymptotically equivalent to the minimum possible risk one would achieve if the true distribution of $\lambda_i\mid Y_{i0}$ were known.  Our analysis builds on the foundational work of \citet{brown2009nonparametric} on compound decision problems, the refinements by \citet{jiang2009general}, and the recent dynamic panel extension of \citet{liu2020forecasting}.

\section{Extensions and tests}\label{sec:extensions-test}
\subsection{Extensions}\label{sec:extensions}
Beyond the baseline specification in \eqref{eq:dyn_panel} and \eqref{eq:delta_ar1}, our proposed method accommodates various extensions to address richer policy questions and realistic data features.  First, one can generalize the treatment indicator $D_{it}^j$ to discrete or continuous dosages $Z_{it}^j$, accommodate staggered adoption designs by allowing treatment timing to vary across units, incorporate time fixed effects $\gamma_t$ further controls for common shocks, and estimate $\delta_{ij}$ for $j \in \{-L, \ldots, -1\}$ to partially check for the no anticipation assumption.

Second, additional covariates $X_{it}$ can be woven into both the QMLE and empirical Bayes steps.  For strictly exogenous controls $X_{it}^O$, their coefficients can be either common or unit-specific, whereas for predetermined covariates $X_{it}^P$, they can only have common coefficients to ensure identification.  These covariate extensions allow researchers to flexibly adjust for observed confounders while still exploiting the shrinkage benefits of empirical Bayes.

Third, the dynamic structure itself can be enriched.  Both the outcome process $Y_{it}$ and the treatment effect sequence $\delta_{ij}$ may follow AR($p$) dynamics; in particular, AR(2) specifications capture potential non-monotonic or oscillatory responses that simple AR(1) models miss.  Moreover, the error term $U_{it}$ can be generalized to admit cross-sectional heteroskedasticity $\sigma_{U,i}^2$ (see for example, \citet{chen2022empirical} and \citet{liu2023density}) or temporal dependence via MA($q$) processes, improving finite sample inference under complex serial correlation patterns.

Finally, our empirical Bayes prior can conditional on various observables: one can consider $\pi(\lambda_i\mid C_i)$, where conditioning variables $C_i$ can include the initial outcome $Y_{i0}$, treatment timing and size $D_{it}^j$ or $Z_{it}^j$, whole time series paths of strictly exogenous covariates $X_{i,0:T}^O$, and initial values of predetermined covariates $X_{i0}^P$.  Under a \emph{conditional strict exogeneity assumption}, namely, the error terms $U_{it}$ is independent of the treatment conditional on $(X_{i,0:T}^O,\,X_{i0}^P)$, these extensions preserve identification and capture richer sources of heterogeneity across units.

\subsection{Tests}\label{sec:tests}
Our analysis not only delivers flexible estimates of treatment effect heterogeneity but also provides a unified toolkit for formally testing model specifications and key event study assumptions.  

In terms of model specification, first, we can examine whether we have random coefficients, where $\lambda_i$ is uncorrelated with $Y_{i0}$, against correlated random coefficients ($H_0\colon b_1=0$).\footnote{Uncorrelation is a necessary but not sufficient condition for independence, making this a more conservative test.} We can test whether there is no correlation between heterogeneous effects and individual heterogeneity, where $\lambda_i$ is uncorrelated with $Y_{i0}$ and $\delta_{ij}$ is uncorrelated with $\alpha_i$ conditional on $Y_{i0}$ ($H_0\colon b_1=0,\,\Sigma_{\lambda,12}=0$). Third, we can check the absence of state dependence in $\delta_{ij}$ ($H_0\colon\rho_{\delta1}=\rho_{\delta2}=0$). See Table \ref{tab:sim-test} for the size and power of these tests in our simulation study, and Table \ref{tab:app-test} for their performance in the county-level recession and unemployment application.

In terms of common event study assumptions, first, as discussed in Remark \ref{rmk:cond-exo}, the parallel trends assumption, such as Assumption 1 in \citet{sun2021estimating}, amounts to zero persistence in $Y_{it}$ absent treatment ($H_0\colon\rho_Y=0$).  Second, the no anticipation assumption, such as Assumption 2 in \citet{sun2021estimating}, requires that $\E[\delta_{ij}]=0$ for $j<0$, which can be tested by verifying that pre-treatment event time coefficients have zero mean. Third, the homogeneous treatment effects assumption, such as Assumption 3 in \citet{sun2021estimating}, implies identical mean treatment paths across cohorts  defined by treatment timing, which can be assessed by comparing the estimated means of $\delta_{ij}$ across these cohorts.  

\section{Monte Carlo simulations}\label{sec:sim}
\subsection{Alternative estimators and DGPs}\label{sec:alt-est-dgp}

\paragraph{Alternative estimators.} In our simulation study, we evaluate two broad groups of estimators for time-varying treatment effects in event studies: the homogeneous treatment effect estimators and the heterogeneous treatment effect ones.  For simplicity, we focus below on the basic setup of Section \ref{sec:model-dynamic} without time fixed effects and additional covariates, and extensions to the generalized model in Section \ref{sec:extensions} can be carried out in a similar manner.

The first group comprises the traditional TWFE without any lagged outcome and an augmented version with an AR(1) term.  The baseline \emph{TWFE} regresses the observed outcome $Y_{it}$ on event time dummies and unit fixed effects,
\[
Y_{it} = \sum_{j=-L}^{J} D_{it}^{j} \delta_{j} + \alpha_i + U_{it},
\]
normalizing the pre-treatment period by setting $\delta_{-1}=0$.  While straightforward, omitting dynamics can lead to omitted variable bias when outcomes are serially correlated. To mitigate this bias, we introduce an augmented \emph{TWFE+AR(1)} estimator, which includes a lagged outcome $Y_{i,t-1}$ as an additional regressor,
\[
Y_{it} = \rho_Y Y_{i,t-1} + \sum_{j=-L}^{J} D_{it}^{j} \delta_{j} + \alpha_i + U_{it},\quad\text{normalizing } \delta_{-1}=0,
\]
while still consider a common effect $\delta_j$ across units.

The second class of estimators allows for unit-specific dynamic responses as in \eqref{eq:dyn_panel}. 
We consider the following four heterogeneous treatment effect estimators, which differ in how they recover the marginal density of the sufficient statistics $p(\widehat\lambda\mid Y_0)$ in Tweedie's formula \eqref{eq:tweedie}.  The \emph{oracle} estimator knows the true distribution and the true common parameters, and thus attains the infeasible optimum to which we benchmark our feasible estimator. The \emph{parametric} estimator adopts a parametric form of the distribution, typically Gaussian, which is in line with the QMLE and easy to implement. The nonparametric estimator models the distribution via \emph{kernel} or \emph{mixture} and offers flexibility to uncover complex heterogeneity patterns at the cost of longer computation time and higher variance.\footnote{For the kernel estimator, we use a Gaussian kernel with bandwidth chosen by Silverman's rule of thumb, which performs well in our simulations and empirical application, although more advanced bandwidth selection methods could further improve its estimation accuracy.}  Our main focus is on the parametric and nonparametric approaches.

\paragraph{DGPs.} We simulate panel data according to a dynamic event study model in \eqref{eq:dyn_panel},
and the treatment effect sequence $\{\delta_{ij}\}_{j=0}^m$ follows an AR($p$) process,
\[
\delta_{ij} = \sum_{k=1}^p\rho_{\delta p} \delta_{i,j-p} + \epsilon_{ij},
\quad \epsilon_{ij}\overset{\mathrm{iid}}{\sim}N(0,\sigma_\epsilon^2),
\]
for $j=p,\dots,J$, with initial draws $\delta_{i0}$.

In our baseline design, we set the cross-sectional sample size to $N=1000$, the time series dimension to $T=10$, the treatment onset to $t_0=5$, and the maximum event horizon to $J=5$.  The common parameters are $\rho_Y=0.8$, $\sigma_U^2=1/T$, and $\sigma_\epsilon^2=1/T$. 

For the distribution of unit-specific parameters $\pi(\lambda \mid Y0)$, we take into account the following four aspects that capture different heterogeneity and state dependence patterns.  First, we explore both normal and non-normal distributions.  Second, we examine both a random coefficients (RC) setup with $\lambda_i\perp Y_{i0}$ and a correlated random coefficients (CRC) setup with $\lambda_i\not\perp Y_{i0}$.  Third, we investigate scenarios where $\alpha_i$ and $\delta_i$ are either independent or correlated conditional on $Y_{i0}$.  Finally, we consider both AR(1) and AR(2) for state dependence in the treatment effect dynamics. For the AR(2), we specify four cases: $(\rho_{\delta,1},\rho_{\delta,2})=$ (0, 0) in Case 1 for no state dependence, (0.3, 0) in Case 2 for pure AR(1), (0.5, 0.2) in Case 3 for a monotonic decay, (0.75, -0.25) in Case 4 for an oscillation response, all with initial means $\E[\delta_{i0}]=3$ and $\E[\delta_{i1}]=1.5$.
For each experimental setup, we execute $N_{\rm sim}=100$ Monte Carlo simulations.  

\subsection{Results}\label{sec:sim-results}
In the main text, we focus on the common parameter estimates, joint distribution of the individual heterogeneity, time-varying treatment effects, and tests, for the specifications with non-normal distribution, correlated random coefficients, $\alpha_i\not\perp\delta_{i}\mid Y_{i0}$, and $\delta_{ij}\sim$ AR(2). For detailed results across all model specifications, please refer to the online appendix. 
The main messages are similar across all specifications.

\begin{table}[t]
\caption{Common parameter estimates by QMLE - Monte Carlo}
\label{tab:sim-common}
\begin{center}
\begin{tabular}{lrrr|rrr} \hline \hline
    & \multicolumn{3}{c|}{Case 1} & \multicolumn{3}{c}{Case 2} \\
	&	Bias	&	SD	&	RMSE	&	Bias	&	SD	&	RMSE	\\\hline
$\rho_Y$	&	0.000	&	0.002	&	0.002	&	0.001	&	0.003	&	0.003	\\
$\rho_{\delta 1}$	&	0.000	&	0.014	&	0.014	&	0.023	&	0.022	&	0.032	\\
$\rho_{\delta 2}$	&	0.000	&	0.008	&	0.008	&	-0.012	&	0.012	&	0.017	\\
$\sigma_U^2$	&	0.000	&	0.003	&	0.003	&	0.000	&	0.003	&	0.003	\\
$\sigma_{\epsilon}^2$	&	-0.001	&	0.005	&	0.005	&	0.003	&	0.005	&	0.006	\\
\hline	
& \multicolumn{3}{c|}{Case 3} & \multicolumn{3}{c}{Case 4} \\												
	&	Bias	&	SD	&	RMSE	&	Bias	&	SD	&	RMSE	\\\hline
$\rho_Y$	&	0.003	&	0.005	&	0.006	&	0.004	&	0.003	&	0.005	\\
$\rho_{\delta 1}$	&	0.037	&	0.024	&	0.043	&	0.027	&	0.016	&	0.031	\\
$\rho_{\delta 2}$	&	-0.028	&	0.014	&	0.031	&	-0.015	&	0.008	&	0.017	\\
$\sigma_U^2$	&	-0.001	&	0.002	&	0.003	&	-0.002	&	0.002	&	0.003	\\
$\sigma_{\epsilon}^2$	&	0.019	&	0.006	&	0.020	&	0.038	&	0.006	&	0.038	\\
\hline
\end{tabular}
\end{center}
{\footnotesize {\em Notes:} DGP: Non-normal, CRC, $\alpha_i\not\perp\delta_{i}\mid Y_{i0}$, $\delta_{ij}\sim$ AR(2). $(\rho_{\delta,1},\rho_{\delta,2})=$ (0, 0) in Case 1, (0.3, 0) in Case 2, (0.5, 0.2) in Case 3, (0.75, -0.25) in Case 4. Initial means: $\E[\delta_{i0}]=3$, $\E[\delta_{i1}]=1.5$.}\setlength{\baselineskip}{4mm}
\end{table}

Table \ref{tab:sim-common} reports the bias, standard error, and RMSE of the QMLE for the common parameters. Standard errors are computed using the robust QMLE variance formula from Theorem \ref{thm:QMLE}. Across all four cases, the QMLE exhibits small bias and variance with RMSE below 0.05 for every parameter.

\begin{figure}[t]
  \caption{Joint distribution of $\widetilde\lambda_i$ - Monte Carlo, random coefficients}
  \label{fig:sim-dist-rc}
  \begin{center}
      \vspace{-1.5em}
  \includegraphics[width=\textwidth]{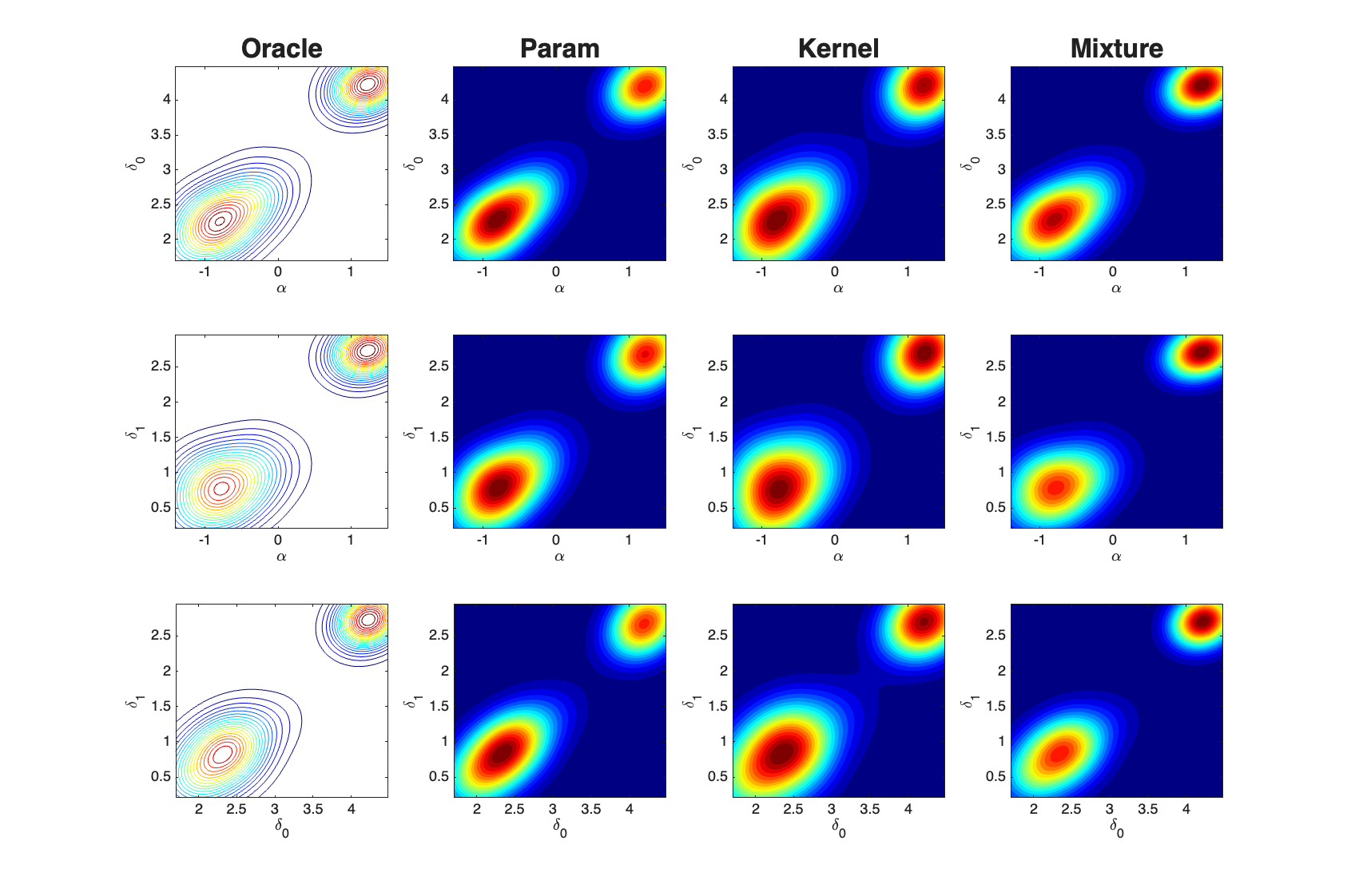}
      \vspace{-2em}
  \end{center}
  {\footnotesize {\em Notes:} DGP: Non-normal, $\alpha_i\not\perp\delta_{i}\mid Y_{i0}$, $\delta_{ij}\sim$ AR(2). Case 3: $(\rho_{\delta,1},\rho_{\delta,2})=$ (0.5, 0.2). Initial means: $\E[\delta_{i0}]=3$, $\E[\delta_{i1}]=1.5$.
  }\setlength{\baselineskip}{4mm}
  \end{figure}

\begin{figure}[t]
  \caption{Joint distribution of $\widetilde\lambda_i$ - Monte Carlo, correlated random coefficients} 
  \label{fig:sim-dist-crc}
  \begin{center}
      \vspace{-1.5em}
  \includegraphics[width=\textwidth]{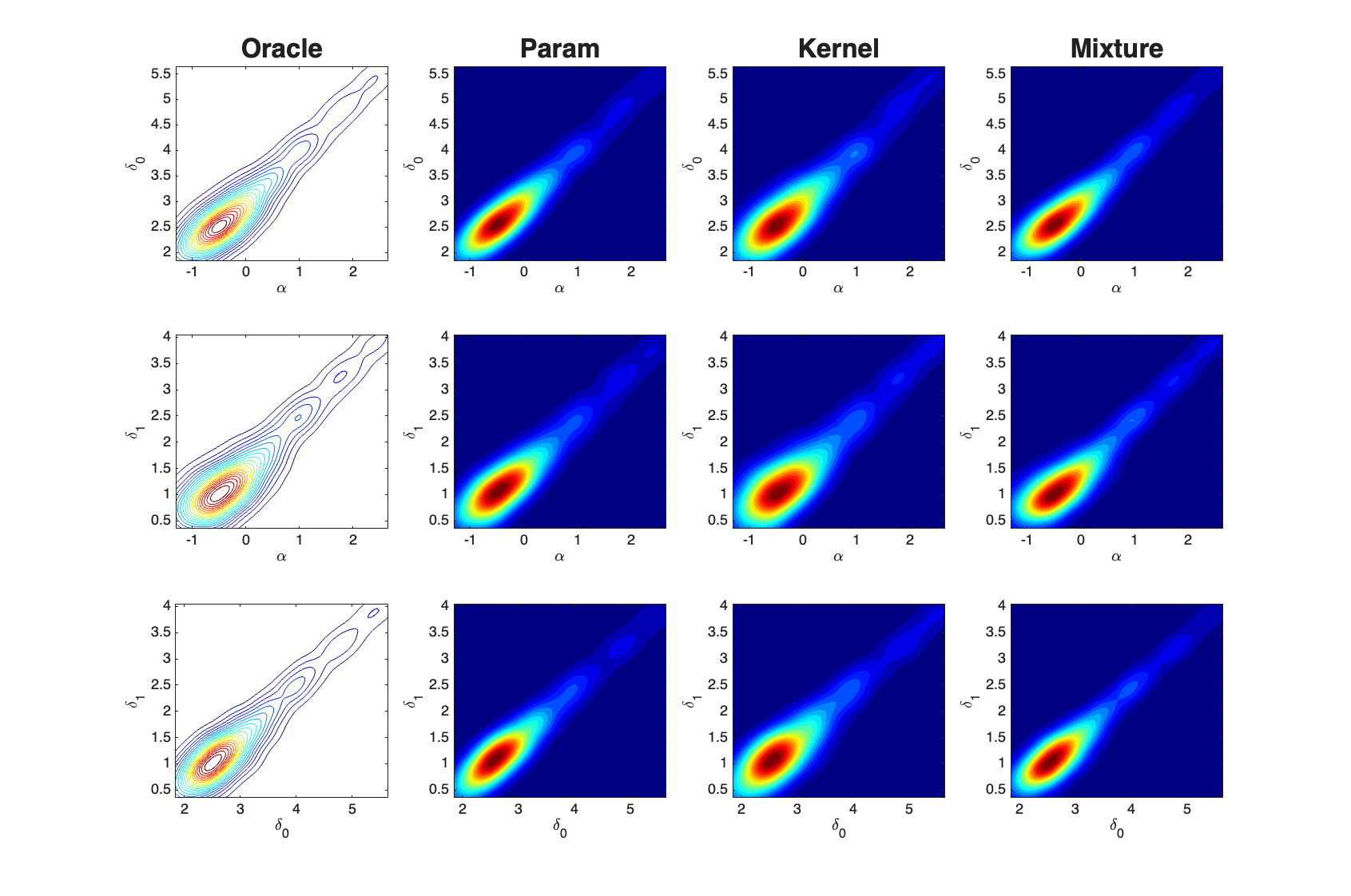}
      \vspace{-2em}
  \end{center}
  {\footnotesize {\em Notes:} DGP: Non-normal, $\alpha_i\not\perp\delta_{i}\mid Y_{i0}$, $\delta_{ij}\sim$ AR(2). Case 3: $(\rho_{\delta,1},\rho_{\delta,2})=$ (0.5, 0.2). Initial means: $\E[\delta_{i0}]=3$, $\E[\delta_{i1}]=1.5$.
  }\setlength{\baselineskip}{4mm}
  \end{figure}

Figures \ref{fig:sim-dist-rc} and \ref{fig:sim-dist-crc} plot the joint distribution of the empirical Bayes estimates $\widetilde\lambda_i=(\widetilde\alpha_i,\widetilde\delta_{i0},\widetilde\delta_{i1})$ via their pairwise marginal heatmaps, for the random coefficients and correlated random coefficients designs, respectively.\footnote{Note that the distribution $p(\widetilde\lambda)$ differs from $\pi(\lambda)$. The former is based on the empirical Bayes posterior means and embeds information from each unit's observed sequence.} The rows correspond to $(\widetilde\alpha_i,\widetilde\delta_{i0})$, $(\widetilde\alpha_i,\widetilde\delta_{i1})$, and $(\widetilde\delta_{i0},\widetilde\delta_{i1})$, from top to bottom, and the columns show the oracle, parametric, kernel, and mixture empirical Bayes estimators, from left to right. All three feasible empirical Bayes estimators produce very similar heatmaps that closely track the oracle benchmark and successfully capture the bimodal pattern in Figure \ref{fig:sim-dist-rc} and the heavy tail behavior in Figure \ref{fig:sim-dist-crc}. Quantitatively, the mixture estimator achieves the lowest RMSE for $\lambda_i$, with roughly a 5--10\% improvement over both the parametric and kernel approaches. The parametric estimator shows a slightly larger bias due to its misspecified Gaussian prior, and the kernel estimator exhibits slightly higher variance due to its nonparametric setup. 

\begin{figure}[t]
  \caption{Event study with time-varying treatment effects - Monte Carlo}
  \label{fig:sim-event-study}
  \begin{center}
      \vspace{-1em}
  \hspace*{-0.1\textwidth}\includegraphics[width=1.2\textwidth]{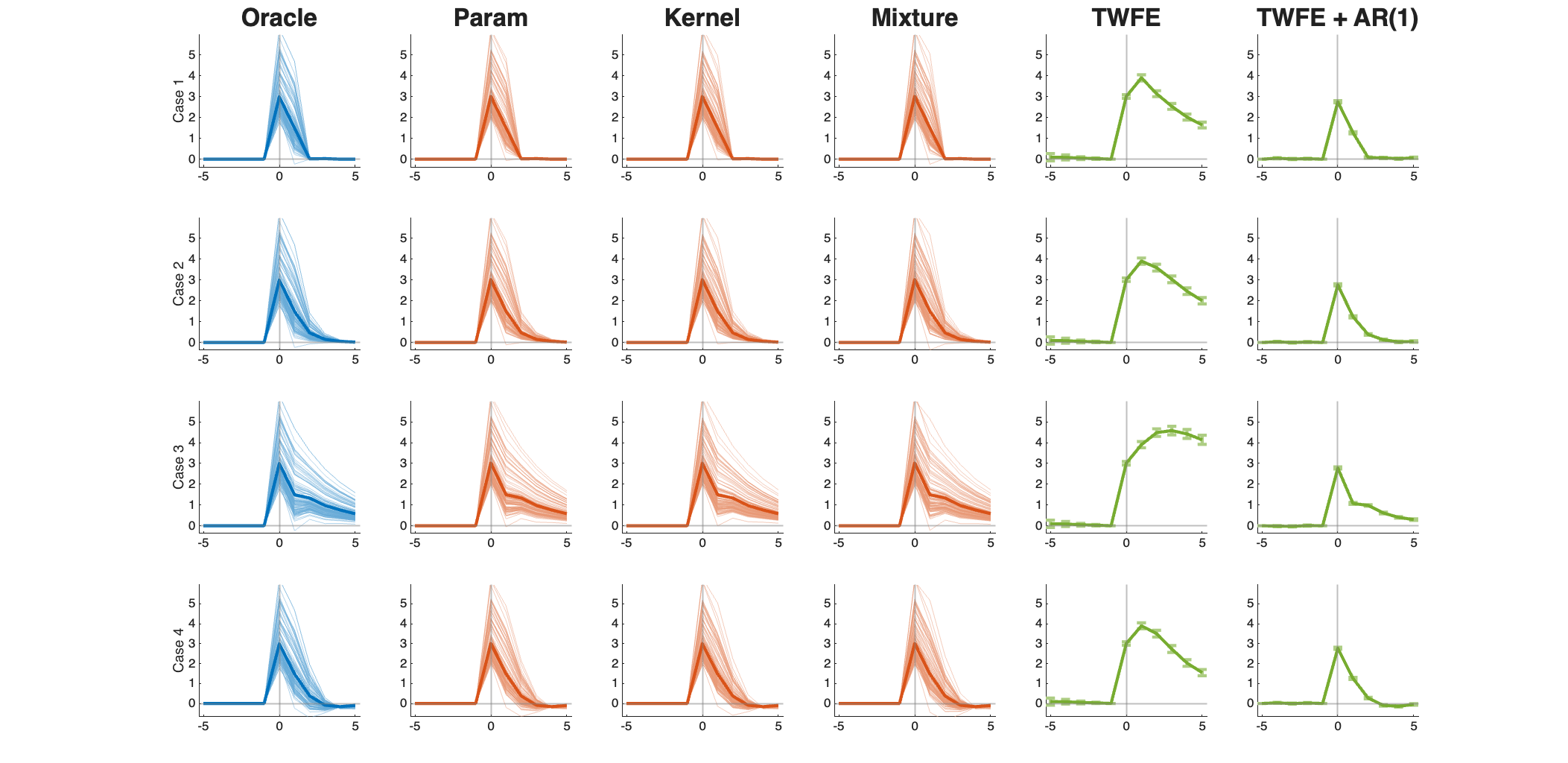}
      \vspace{-1.5em}
  \end{center}
  {\footnotesize {\em Notes:} DGP: Non-normal, CRC, $\alpha_i\not\perp\delta_{i}\mid Y_{i0}$, $\delta_{ij}\sim$ AR(2). $(\rho_{\delta,1},\rho_{\delta,2})=$ (0, 0) in Case 1, (0.3, 0) in Case 2, (0.5, 0.2) in Case 3, (0.75, -0.25) in Case 4. $\E[\delta_{i0}]=3$, $\E[\delta_{i1}]=1.5$. TWFE and TWFE+AR(1): bars indicate 95\% CI, clustered s.e. by unit.
  }\setlength{\baselineskip}{4mm}
  \end{figure}

Figure \ref{fig:sim-event-study} displays the estimated heterogeneous dynamic treatment effect paths across event time for four DGP scenarios.  From top to bottom, the rows show Cases 1--4: no state dependence, pure AR(1), monotonic AR(2), and oscillatory AR(2). From left to right, the columns present the infeasible optimum oracle estimator, followed by the parametric, kernel, and mixture empirical Bayes estimators, as well as the homogeneous TWFE and TWFE+AR(1) estimators. In each graph, the thin lines depict the heterogeneous dynamic responses of the individual units. As before, all feasible empirical Bayes estimators yield trajectories nearly indistinguishable from the oracle benchmark and accurately recover each DGP's dynamic patterns, whether simple exponential decay, gradual tapering, or sign-changing oscillation, thereby recovering substantial dynamic heterogeneity across units.

The last two columns are the homogeneous estimators. The baseline TWFE estimator fails to account for the dynamics in the outcome, and produces substantial misspecification bias with larger and more persistent estimated effects.  For the augmented TWFE+AR(1), its estimated mean path aligns closely with the true mean pattern, but it is not able to capture the cross-unit dispersion, and its 95\% confidence bands are too narrow to reflect underlying heterogeneity.  In contrast, our empirical Bayes estimators efficiently combine information across all units and flexibly adapt to each unit's own response profile, and thus deliver good estimates of the average treatment path and effectively capture the heterogeneity in dynamics.

\begin{table}[t]
\caption{Rejection rates of tests - Monte Carlo}
\label{tab:sim-test}
\begin{center}
\begin{tabular}{lrrrr|rrrr|rrrr} \hline \hline
    & \multicolumn{4}{c|}{RC, $\alpha_i\perp\delta_{i}\mid Y_{i0}$} & \multicolumn{4}{c|}{RC, $\alpha_i\not\perp\delta_{i}\mid Y_{i0}$} & \multicolumn{4}{c}{CRC, $\alpha_i\not\perp\delta_{i}\mid Y_{i0}$} \\
    Case &	1	&	2	&	3	&	4	&	1	&	2	&	3	&	4 &	 1	&  2	&	3	&	4	\\\hline
    Test 1 &	\textcolor{blue}{0.04}	&	\textcolor{blue}{0.04}	&	\textcolor{blue}{0.04}	&	\textcolor{blue}{0.06}	&	\textcolor{blue}{0.06}	&	\textcolor{blue}{0.06}	&	\textcolor{blue}{0.05}	&	\textcolor{blue}{0.04}	&	1.00	&	1.00	&	1.00	&	1.00	\\
    Test 2 &	\textcolor{blue}{0.05}	&	\textcolor{blue}{0.06}	&	\textcolor{blue}{0.04}	&	\textcolor{blue}{0.06}	&	1.00	&	1.00	&	1.00	&	1.00	&	1.00	&	1.00	&	1.00	&	1.00	\\
    Test 3 &	\textcolor{blue}{0.04}	&	1.00	&	1.00	&	1.00	&	\textcolor{blue}{0.03}	&	1.00	&	1.00	&	1.00	&	\textcolor{blue}{0.02}	&	1.00	&	1.00	&	1.00	\\
\hline	
\end{tabular} 
\end{center}
{\footnotesize {\em Notes:} DGP: Non-normal, $\delta_{ij}\sim$ AR(2). $(\rho_{\delta,1},\rho_{\delta,2})=$ (0, 0) in Case 1, (0.3, 0) in Case 2, (0.5, 0.2) in Case 3, (0.75, -0.25) in Case 4. Initial means: $\E[\delta_{i0}]=3$, $\E[\delta_{i1}]=1.5$. Blue entries: size; black entries: power. Based on robust s.e.}\setlength{\baselineskip}{4mm}
\end{table}

Table \ref{tab:sim-test} reports the rejection rates over 100 simulations for three tests regarding the heterogeneity pattern. As described in Section \ref{sec:tests}, Test 1 checks for random versus correlated random coefficients, Test 2 for the joint independence of $\delta_{ij}$ against $(\alpha_i,Y_{i0})$, and Test 3 for the state dependence in the treatment effect processes. 

The table is partitioned into three blocks. The left block reports rejection rates under a random coefficients DGP in which $\alpha_i\perp\delta_{i}\mid Y_{i0}$, satisfying the null hypotheses of Tests 1 and 2. The middle block corresponds to a random coefficients DGP with $\alpha_i\not\perp\delta_{i}\mid Y_{i0}$, which satisfies Test 1's null but violates Test 2's. The right block is based on a correlated random coefficients DGP with $\alpha_i\not\perp\delta_{i}\mid Y_{i0}$, violating the nulls of both Tests 1 and 2. Within each block, columns give results for Cases 1--4: no AR, AR(1), monotonic AR(2), oscillatory AR(2), where Case 1 conforms to Test 3's null and Cases 2--4 lie under its alternative. 

Together, the blue entries indicate the size of the tests, while the black entries show their power.
Under the null hypotheses,  all tests maintain size close to the nominal 5 \% level, with rejection rates between 0.04 and 0.06.\footnote{One observed size of 0.02 likely reflects Monte Carlo noise with only 100 repetitions.}  Under the alternative hypotheses, the power is 1.00, possibly due to the relatively large sample size with $N=1000$ and $T=10$. Therefore, these tests provide a reliable means of diagnosing the heterogeneity pattern and state dependence structure.  In particular, these tests allow us to assess whether treatment effect dynamics are driven primarily by unobserved baseline heterogeneity or by the initial treatment impact.


\section{Empirical example: recession and unemployment}\label{sec:app}

\subsection{Data and sample}\label{sec:app-data}
Understanding how recessions shape local labor markets is crucial for designing targeted policy responses. The 2008 Great Recession led to a nationwide spike in unemployment, peaking at nearly 10\% in October 2009, and ushered in a protracted recovery that saw the national rate fall back to pre-crisis levels only by late 2015.\footnote{See the BLS website, such as \url{https://www.bls.gov/spotlight/2012/recession/pdf/recession_bls_spotlight.pdf} and \url{https://www.bls.gov/news.release/archives/empsit_01082016.pdf}} However, aggregate figures mask substantial variation across regions: some counties experienced sharp spikes, while others bore delayed and milder losses.
For example, \citet{yagan2019employment} documents long-lasting employment and earnings losses for harder-hit areas, and \citet{hershbein2020recessions} further show that those areas also experienced persistent population declines.

In this empirical example, we exploit county-level unemployment data to map these heterogeneous responses over time. Our outcome, $Y_{it}$, is the annual unemployment rate for county $i$ in year $t$. We define the onset of the Great Recession as 2008, assigning it to period $t_0=5$ within a ten year window. The sample spans 2003--2013 ($T=10$) across $N=3142$ U.S. counties, capturing five pre- and five post-recession years.  The county-level not seasonally adjusted unemployment rates are obtained from the Bureau of Labor Statistics (BLS) website, and we aggregate the monthly data to an annual frequency by time averaging. 

This panel event study analysis allows us to estimate county-specific dynamic effects while controlling for unobserved heterogeneity and serial dependence, thereby shedding light on both the immediate and persistent impacts of the recession across diverse local economies.

\subsection{Results}\label{sec:app-results}
In this section, we focus on the estimator under the AR(2) specification for $\delta_{ij}$. Analogous results for the AR(1) case and models with time fixed effects are provided in the online appendix. 

\begin{table}[t]
\caption{Common parameter estimates by QMLE - recession and unemployment example}
\label{tab:app-common}
\begin{center}
\begin{tabular}{lrr|lrr} \hline \hline
        &	Est.\	&	SD\textcolor{white}{x}	&	&	Est.\	&	SD\textcolor{white}{x}	\\\hline
    $\rho_Y$	&	\textcolor{black}{0.845}	&	(0.010)	&	$\sigma_U^2$	&	0.431	&	(0.103)	\\
    $\rho_{\delta 1}$	&	\textcolor{black}{0.306}	&	(0.011)	&	$\sigma_{\epsilon}^2$	&	0.276	&	(0.094)		\\
    $\rho_{\delta 2}$	&	-0.061	&	(0.011)	&		&		&		\\
    \hline
    \end{tabular} 
\end{center}
\end{table}

In Table \ref{tab:app-common} for common parameter estimates, the estimated persistence in the unemployment rate is high and significant with $\widehat\rho_Y=0.845$, so the omitted variable bias could be substantial for the traditional TWFE regression.  The AR(2) dynamics of the recessionary effect are likewise significant with $\widehat\rho_{\delta1}=0.306$ and $\widehat\rho_{\delta2}=-0.061$, indicating a damped oscillatory decay in local labor market responses.

\begin{figure}[t]
  \caption{Joint distribution of $\widetilde\lambda_i$ - recession and unemployment example} 
  \label{fig:app-dist}
  \begin{center}
      \vspace{-1.5em}
  \includegraphics[width=.8\textwidth]{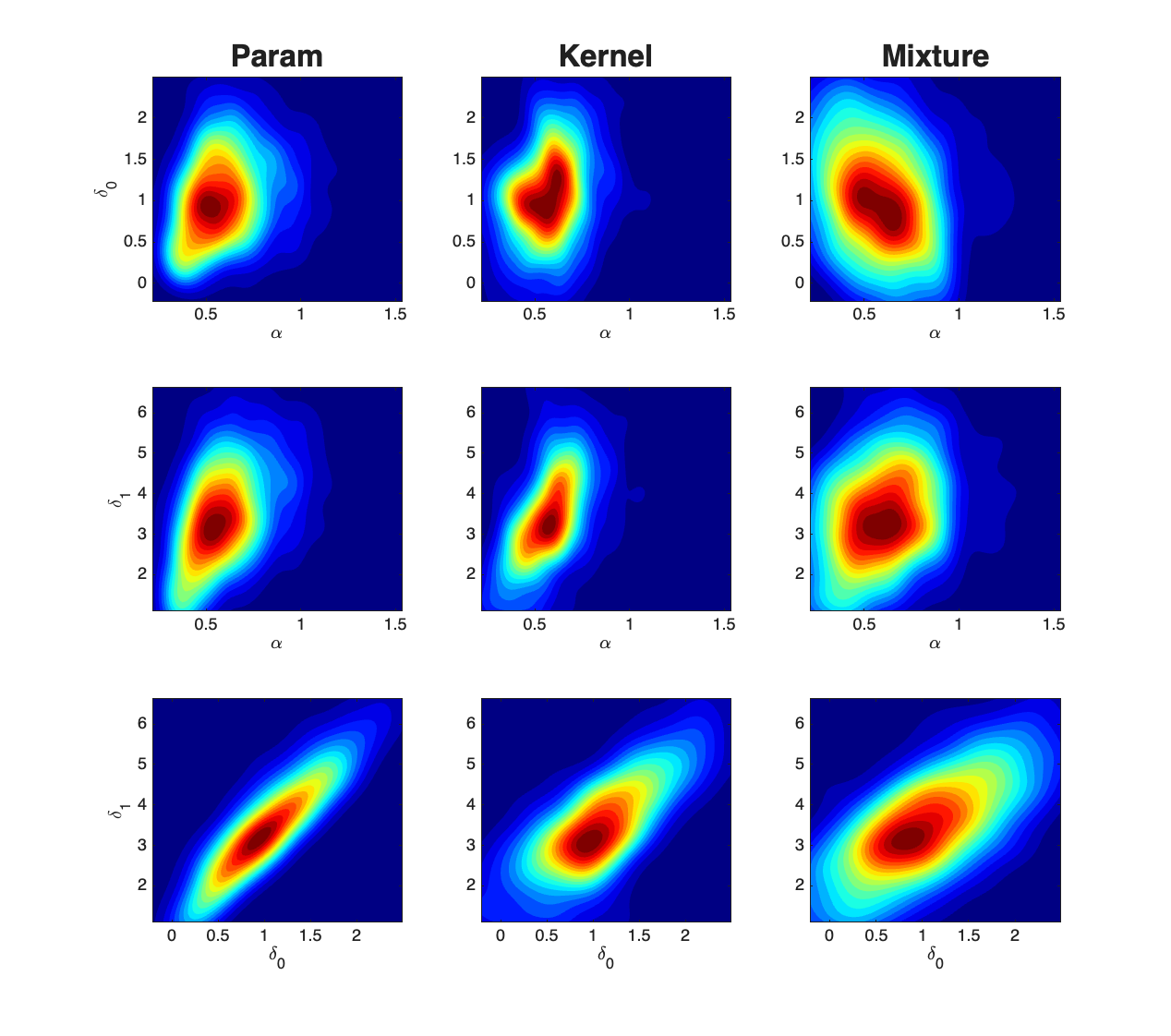}
      \vspace{-2em}
  \end{center}
  \end{figure}

  Figure \ref{fig:app-dist} presents the heatmaps of the joint distributions of empirical Bayes posterior means across the parametric, kernel, and mixture estimators. All three estimators yield qualitatively similar density shapes. The heatmaps also reveal strong non-Gaussian heterogeneity with asymmetric mass and possible heavy tails rather than simple elliptical contours.  In the first two rows, counties with higher baseline heterogeneity $\alpha_i$ tend to exhibit larger initial effects $(\delta_{i0},\delta_{i1})$, indicating that areas already suffering from high unemployment were hit hardest by the recession.  The third row shows a strong positive correlation between $\delta_{i0}$ and $\delta_{i1}$, reflecting persistent temporal dynamics in treatment responses.  These irregular patterns underscore the value of the flexible empirical Bayes methods for jointly modeling $(\alpha_i,\delta_{i0},\delta_{i1})$ and uncovering the rich heterogeneity across counties.

\begin{figure}[t]
  \caption{Event study w.\ time-varying treatment effects - recession \& unemployment example}
  \label{fig:app-event-study}
  \begin{center}
      \vspace{-1em}
  \hspace*{-0.1\textwidth}\includegraphics[width=1.2\textwidth]{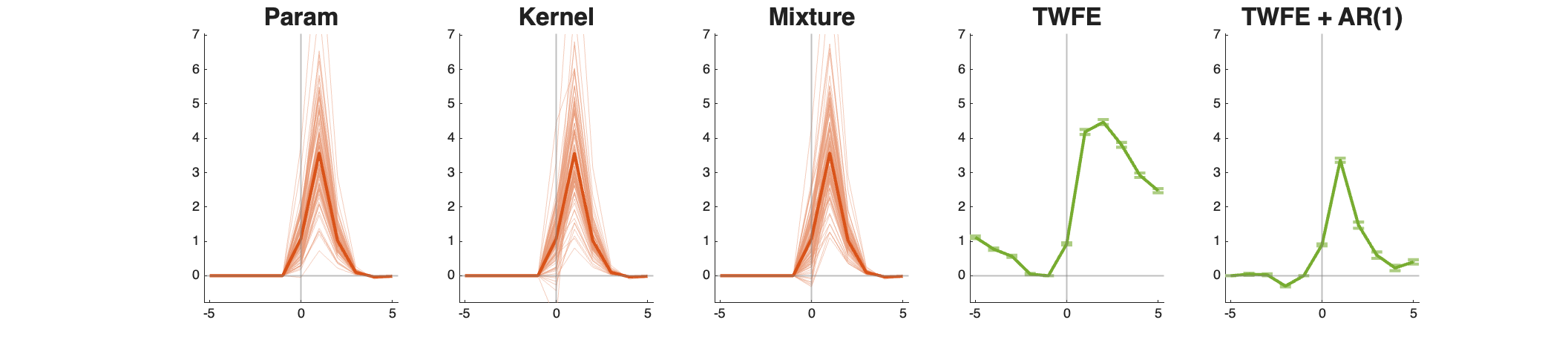}
      \vspace{-1.5em}
  \end{center}
  {\footnotesize {\em Notes:} TWFE and TWFE+AR(1): bars indicate 95\% CI, clustered s.e. by unit.
  }\setlength{\baselineskip}{4mm}
  \end{figure}

  Figure \ref{fig:app-event-study} plots county-specific event study estimates of the time-varying treatment effects.  As seen in the joint distributions, all three empirical Bayes estimators produce qualitatively similar trajectories.  The individual curves reveal stark heterogeneity: some counties suffered a dramatic jump in unemployment of over 7 percentage points in 2009, others experienced only modest rises of around 0.5 points, and a few even registered slight declines in the initial recession year 2008.  These spikes and the varied post-2008 decay profiles far exceed the average effect implied by the TWFE model. In particular, the baseline TWFE yields pre-2008 coefficients that are significantly different from zero, indicating substantial omitted variable bias from ignoring the serial dependence of unemployment.

\begin{table}[t]
\caption{Tests  - recession and unemployment example}
\label{tab:app-test}
\begin{center}
\begin{tabular}{lrrr} \hline \hline
     &	Test stat	&	Crit.\ val.\	&	Reject?		\\\hline
    Test 1 &	672.6	&	5.99	&	Y	\\
    Test 2 &	766.7	&	9.49	&	Y	\\
    Test 3 &	1069.2	&	5.99	&	Y	\\
\hline	
\end{tabular}  
\end{center}
{\footnotesize {\em Notes:} Based on QMLE estimates with robust s.e. Test 1: $H_0\colon b_1=0$; Test 2: $H_0\colon b_1=0,\,\Sigma_{\lambda,12}=0$; Test 3: $H_0\colon\rho_{\delta1}=\rho_{\delta2}=0$. Critical values: 5\% level.
  }\setlength{\baselineskip}{4mm}
\end{table}

Finally, Table \ref{tab:app-test} formally tests three key modeling assumptions: see Section \ref{sec:tests} for a more detailed description of the tests. The rejections of all three tests reveal several key features of the Great Recession's impact on U.S.\ local labor markets.  First, rejecting the pure random coefficients null (Test 1) shows that the unobserved heterogeneity, including the treatment effects, is not idiosyncratic but instead systematically related to county characteristics: places with higher pre-crisis unemployment were hit especially hard.  Second, the rejection of the joint independence null (Test 2) confirms a strong link between baseline heterogeneity and dynamic responses, indicating that local labor market resilience or vulnerability cannot be treated as exogenous.  Finally, ruling out the no state dependence null (Test 3) demonstrates that the recessionary impact on local labor markets is not a one-off hit but unfolds dynamically, with early effects shaping subsequent recovery or further distress. Together, these results highlight the inadequacy of homogeneous static TWFE specifications and validate the need for our dynamic heterogeneous panel framework.


\section{Conclusion}\label{sec:conclusion}
In summary, our paper makes three key contributions. First, we demonstrate how omitting predetermined variables can severely bias event study estimates, and we introduce a semiparametric dynamic panel model with correlated random coefficients that simultaneously captures outcome persistence and treatment effect heterogeneity. Second, we develop a two-step estimator---QMLE for common parameters followed by an empirical Bayes correction for unit-specific effects---that is easy to implement and achieves oracle risk performance. Finally, our analysis offers new insights into standard event study assumptions, including no anticipation, homogeneous treatment effects across treatment timing cohorts, and state dependence structure, making it easier to diagnose and address potential violations in empirical research.

The potential applications of our method extend to any setting with short panel data where we are interested in the dynamics of the heterogeneous treatment effects. In corporate finance, it can revisit classic event studies of earnings announcements, mergers, or regulatory changes, allowing for firm-level persistence and heterogeneous responses. In public policy, it can evaluate staggered social program roll-outs, uncovering differential impacts across communities or demographic groups. Likewise, research in health, education, environmental policy, labor markets, and macroprudential regulation can potentially benefit by using our semiparametric, shrinkage-based estimator to produce more accurate estimates of how treatment effects evolve over time.

\clearpage
\begin{singlespace}
    \bibliographystyle{econometrica}
\bibliography{TVTE_lit}
\end{singlespace}
\clearpage

\begin{center}
{\Large Appendix:\\ \vspace{0.5em} Time-Varying Heterogeneous Treatment Effects in Event Studies}
\vspace{1.5em}

{\large Irene Botosaru  \hspace{1em} Laura Liu}
\vspace{0.5em}

{\large\today}
\end{center}
\vspace{0.5em}

\appendix
\renewcommand{\theequation}{A.\arabic{equation}}
\setcounter{equation}{0} 
\renewcommand{\thefigure}{A.\arabic{figure}}
\setcounter{figure}{0} 
\renewcommand{\thetable}{A.\arabic{table}}
\setcounter{table}{0} 
\setcounter{footnote}{0} 
\setcounter{page}{1}
\renewcommand{\thepage}{A-\arabic{page}}


\section{Proofs}\label{sec:proofs}
\subsection{Identification}
\begin{proof}[Proof of Theorem \ref{thm:ident}]
We prove identification in two steps, building on the approach of \citet{arellano2012identifying}. First, we establish identification of the parameters $\rho=(\rho_Y, \rho_\delta)'$.
Second, given identified $\rho$, we show that the conditional density $\pi(\lambda_i \mid Y_{i0})$ is identified via characteristic function deconvolution.
\paragraph{Step 1: Identification of common parameters $\rho$.} 

Under Assumption \ref{assu:model} for model setup, we identify $\rho$ via the following moment conditions.

First, for the autoregressive parameter $\rho_Y$, under Assumption \ref{assu:rank}, $t_0 \geq 3$ provides at least two pre-treatment periods, and the moment condition for $\rho_Y$ is
\[
\E\left[\frac{1}{N} \sum_{i=1}^N \sum_{t=1}^{t_0-1} (Y_{it} - \rho_Y Y_{i,t-1} - \overline{Y}_{t} + \rho_Y \overline{Y}_{t-1}) Y_{i,t-1}\right] = 0,
\]
where $\overline{Y}_t = N^{-1} \sum_{i=1}^N Y_{it}$ helps remove the individual levels $\alpha_i$. Assumption \ref{assu:dist}(c) ensures $\Var(Y_{i0}) > 0$, and thus this moment condition is non-degenerate.

Second, for treatment effect persistence $\rho_\delta$, using treatment and post-treatment periods $t \geq t_0$, we exploit the autoregressive structure of $\delta_{ij}$. Let $\widetilde{Y}_{it} = Y_{it} - \rho_Y Y_{i,t-1}$ denote the transformed outcome. The moment condition is:
\[
\E\left[\frac{1}{N} \sum_{i=1}^N \sum_{t=t_0+1}^T \widetilde{Y}_{it} \widetilde{Y}_{i,t-1} \right] = \rho_\delta \E\left[\frac{1}{N} \sum_{i=1}^N \sum_{t=t_0+1}^T \widetilde{Y}_{i,t-1}^2 \right].
\]
Under Assumption \ref{assu:rank}, the condition $T - t_0 \geq J \geq 1$ ensures sufficient post-treatment observations. Moreover, Assumption \ref{assu:dist}(c) ensures $\Var(\delta_{i0}) > 0$, and thus this moment condition is non-degenerate.

\paragraph{Step 2: Identification of $\pi(\lambda_i \mid Y_{i0})$ given identified $\rho$.}

Having identified $\rho$ in Step 1, we now verify the conditions of Theorem 2 in \citet{arellano2012identifying} for deconvolving the conditional density $\pi(\lambda_i \mid Y_{i0})$. The true composite error $\check U_{i,1:T}(\rho_{\delta,0})$ has the MA($J$) structure
\[
\check U_{it}(\rho_{\delta,0}) = U_{it} + \sum_{j=0}^{J} \sum_{k=1}^j \rho_{\delta,0}^{j-k} D_{it}^j \varepsilon_{ik}.
\]

First, for their Assumption 1 (Mean independence) and Assumption 3 (Conditional independence), our simple model with common treatment timing $t_0$ together with Assumption \ref{assu:model}(b) ensures that $\E[\check U_{it}(\rho_{\delta,0}) \mid \lambda_i, Y_{i0}] = 0$ and $\check U_{it}(\rho_{\delta,0}) \perp \lambda_i \mid Y_{i0}$. Since $W_i(\rho_{\delta,0})$ is deterministic and identical across units, we omit it from the conditioning set.

Second, for their Assumption 4 (Non-vanishing characteristic functions), our Assumption \ref{assu:dist}(a) directly imposes that the characteristic functions of $\lambda_i \mid Y_{i0}$, $U_{it}$, and $\varepsilon_{ij}$ are non-vanishing almost everywhere, which extends to $\check U_{it}(\rho_0)$.

Third, for their Assumption 5 (MA structure), the key insight is that our composite error involves exactly $m = 2$ fundamental variance components from $U_{it}$ and $\varepsilon_{ij}$, given the model structure in \eqref{eq:dyn_panel} and \eqref{eq:delta_ar1}. Following from Assumption \ref{assu:dist}(a,b), the hessian of the log characteristic function of $\check U_{it}(\rho_0)$ exists almost everywhere.
The hessian can be decomposed as
\[
\mathrm{vec}\left(\frac{\partial^2 \log \Psi_{\check U_{i,1:T}(\rho_0)}(\tau)}{\partial \tau \partial \tau'}\right) = \mathcal{S} \omega(\tau),
\] for $\tau \in \mathbb{R}^T$,
where $\omega(\tau) = (\omega_U(\tau), \omega_{\varepsilon}(\tau))'$ with
\[
\omega_U(\tau) = \frac{\partial^2 \log \Psi_{U}(\tau)}{\partial \tau^2}, \quad
\omega_{\varepsilon}(\tau) = \frac{\partial^2 \log \Psi_{\varepsilon}(\tau)}{\partial \tau^2}.
\]
The selection matrix $\mathcal S=S(\{D_{it}^j\}, \rho_{\delta,0})$ encodes the treatment pattern and MA lag structure.

Fourth, for their rank condition in equation (24), $\mathrm{rank}(M_i \mathcal S) = m=2,$ where $M_i=\I_{T^2}-(W_i \otimes W_i)[(W_i \otimes W_i)'(W_i \otimes W_i)]^{-1}(W_i \otimes W_i)'$ projects out the design matrix effect. Here we suppress the dependence on $\rho_0$ for notational simplicity. To illustrate, consider the minimal case $T=4$, $t_0=3$, $J=1$. The variance-covariance matrix of $\check U_{i,1:4}$ is
\[
\Sigma_{\check U}= \begin{pmatrix}
\sigma_{U}^2 & 0 & 0 & 0 \\
0 & \sigma_{U}^2 & 0 & 0 \\
0 & 0 & \sigma_{U}^2 & 0 \\
0 & 0 & 0 & \sigma_{U}^2 + \sigma_{\varepsilon}^2
\end{pmatrix},
\]
and the selection matrix $\mathcal S$ is $16 \times 2$ and encodes how $(\sigma_U^2, \sigma_\varepsilon^2)$ contribute to $\mathrm{vec}(\Sigma_{\check U})$.
The design matrix is \[W_i = \begin{pmatrix} 1 & 0 \\ 1 & 0 \\ 1 & 1 \\ 1 & 1+\rho_{\delta,0} \end{pmatrix}.\] One can verify that $\mathrm{rank}(M_i \mathcal S) = 2$ and satisfying the identification condition.

More generally, under Assumption \ref{assu:rank}, $t_0 \geq 3$ provides sufficient pre-treatment and treatment periods to satisfy the degrees of freedom bound $m = 2 \leq \frac{t_0(t_0+1)}{2} - \frac{d_{\lambda}(d_{\lambda}+1)}{2}$ where $d_{\lambda} = 2$: see Remark 3 and equation (27) in \citet{arellano2012identifying}. Then, the projection matrix $M_i$ removes the variation attributable to the heterogeneous parameters $\lambda_i$, leaving sufficient variation from the two variance components $(\sigma_U^2, \sigma_\varepsilon^2)$ to achieve identification, and the rank condition $\mathrm{rank}(M_i \mathcal S) = 2$ holds.

Unlike standard applications, our design matrix $W_i(\rho)$ depends on unknown $\rho$. Our two-step approach resolves this because the identification of $\rho$ in Step 1 uses only the covariance structure of the data and does not require knowledge of $\pi(\lambda_i \mid Y_{i0})$.

Finally, we have the sufficient statistic representation
\[
\widehat\lambda_i(\rho_0) = W_i(\rho_0)^+ (Y_{i,1:T} - \rho_{Y,0} Y_{i,0:T-1}) = \lambda_i + V_i(\rho_0),
\]
where $V_i(\rho_0) = W_i(\rho_0)^+ \check U_{i,1:T}(\rho_0)$ is the projection noise. Since the conditions of Theorem 2 in \citet{arellano2012identifying} have been verified above, characteristic function deconvolution yields
\[
\Psi_{\lambda_i \mid Y_{i0}}(\tau \mid Y_{i0}) = \frac{\Psi_{\widehat\lambda_i(\rho_0) \mid Y_{i0}}(\tau \mid Y_{i0})}{\Psi_{V_i(\rho_0)}(\tau)},
\] for $\tau \in \mathbb{R}^2$,
and the conditional density is recovered via inverse Fourier transform.
\hfill\end{proof}

\subsection{QMLE}
\noindent\begin{proof}[Proof of Theorem \ref{thm:QMLE}] As defined in the main text, $\theta=(\rho_Y,\rho_\delta,\sigma_U^2,\sigma_\varepsilon^2)'$ denotes the common parameters, and $\eta=(\theta',b_0',b_1',\text{vech}(\Sigma_\lambda)')'$ collects both the common parameters and Gaussian random effects parameters.
    Recall that the marginal quasi-log-likelihood is
    \begin{align}
    \ell_N(\eta)
    =-\frac{N}{2}\log\left|\Omega(\eta)\right|
    -\frac{1}{2}\sum_{i=1}^N
    \left(Y_{i,1:T}-\mu_i(\eta)\right)'
    \Omega(\eta)^{-1}
    \left(Y_{i,1:T}-\mu_i(\eta)\right),\label{eq:qll}
    \end{align}
    where
    \begin{align*}
    \mu_i(\eta)&=\mu_i(\rho_Y,\rho_\delta,b_0,b_1)
    =A(\rho_Y) Y_{i0}+\widetilde{W}(\rho_Y,\rho_\delta)(b_0+b_1Y_{i0}),\\
    \Omega(\eta)&=\Omega(\rho_Y,\rho_\delta,\sigma_U^2,\sigma_\varepsilon^2,\Sigma_\lambda)
    =B(\rho_Y)\Sigma_{\check U}(\rho_\delta,\sigma_U^2,\sigma_\varepsilon^2)B(\rho_Y)'
    +\widetilde{W}(\rho_Y,\rho_\delta)\Sigma_\lambda \widetilde{W}(\rho_Y,\rho_\delta)',
    \end{align*}
    and \begin{align*}
        A(\rho_Y) &= (\rho_Y, \rho_Y^2, \rho_Y^3, \cdots, 
        \rho_Y^{T})',\\
        B(\rho_Y) &= \begin{pmatrix}
        1 & 0 & 0 & \cdots & 0 \\
        \rho_Y & 1 & 0 & \cdots & 0 \\
        \rho_Y^2 & \rho_Y & 1 & \cdots & 0 \\
        \vdots & \vdots & \vdots & \ddots & \vdots \\
        \rho_Y^{T-1} & \rho_Y^{T-2} & \rho_Y^{T-3} & \cdots & 1
        \end{pmatrix} ,\\
        \widetilde{W}(\rho_Y,\rho_\delta) &= B(\rho_Y)W
        (\rho_\delta).
        \end{align*}

Let $s=\partial \ell_N/\partial\eta$ denote the score. We now show that under correct conditional mean and covariance, the QMLE satisfies
\(
\E\left[s(\eta_0)\mid Y_{i0}\right] = 0
\)
at the true parameter values.
    
\noindent\textbf{(i) Random effects mean parameters $b_0$ and $b_1$.} These derivatives only involve the mean.
    \begin{align*}
    s_{b_0}&=\frac{\partial \ell_N}{\partial b_0}
    =\sum_{i=1}^N 
    \widetilde{W}(\rho_Y,\rho_\delta)'\Omega(\eta)^{-1}(Y_{i,1:T}-\mu_i(\eta)),\\
    s_{b_1}&=\frac{\partial \ell_N}{\partial b_1}
    =\sum_{i=1}^{N}\widetilde{W}(\rho_Y,\rho_\delta)'\Omega(\eta)^{-1}(Y_{i,1:T}-\mu_i(\eta))Y_{i0}.
    \end{align*}
    Since \(\E[Y_{i,1:T}-\mu_i(\eta_0) \mid Y_{i0}]=0\),
    we have \(\E[s_{b_0}(\eta_0)\mid Y_{i0}]=0\) and \(\E[s_{b_1}(\eta_0)  \mid Y_{i0}]=0\).
    
\noindent\textbf{(ii) Covariance parameters \(\theta_\sigma=(\sigma_U^2,\sigma_\varepsilon^2,\text{vech}(\Sigma_\lambda)')'\).} These derivatives only involve the covariance matrix.
There are five parameters in \(\theta_\sigma\). For $k=1,\dots,5$,
    \begin{align*}
    s_{\theta_{\sigma,k}}&=\frac{\partial \ell_N}{\partial \theta_{\sigma,k}}\\
    &=-\frac{N}{2}\tr\left[\Omega(\eta)^{-1}\frac{\partial \Omega}{\partial \theta_{\sigma,k}}(\eta)\right]
    +\frac{1}{2}\sum_{i=1}^N
    (Y_{i,1:T}-\mu_i(\eta))'\Omega(\eta)^{-1}\frac{\partial \Omega}{\partial \theta_{\sigma,k}}(\eta)\Omega(\eta)^{-1}(Y_{i,1:T}-\mu_i(\eta)).
    \end{align*}
As \(\E[x'Ax]=\tr(A\Var(x))\) for \(x\sim(0,\Var(x))\) and \(\Var(Y_{i,1:T}-\mu_i(\eta_0)\mid Y_{i0})=\Omega(\eta_0)\),
    the second term cancels out the first term, and we have \(\E[s_{\theta_{\sigma,k}}(\eta_0)\mid Y_{i0}]=0\).

\noindent\textbf{(iii) Dynamic parameters $\rho_\delta$ and $\rho_Y$.} These derivatives combine both the mean and covariance matrix. For $\rho_k \in \{\rho_\delta, \rho_Y\}$,
\begin{align*}
s_{\rho_k}=\frac{\partial \ell_N}{\partial \rho_k}
&=\underbrace{\sum_{i=1}^N \frac{\partial \widetilde{W}(\rho_Y,\rho_\delta)}{\partial \rho_k}\Omega(\eta)^{-1}\left(Y_{i,1:T}-\mu_i(\eta)\right)(b_0+b_1Y_{i0})}_{(1)}+\underbrace{-\frac{N}{2}\tr\left[\Omega(\eta)^{-1}\frac{\partial \Omega}{\partial \rho_k}(\eta)\right]}_{(2)}\\
&\quad+\underbrace{\frac{1}{2}\sum_{i=1}^N(Y_{i,1:T}-\mu_i(\eta))'\Omega(\eta)^{-1}\frac{\partial \Omega}{\partial \rho_k}(\eta)\Omega(\eta)^{-1}(Y_{i,1:T}-\mu_i(\eta))}_{(3)},
\end{align*}
where the (1) is from the mean and $\E[(1)\mid Y_{i0}]=0$ by a similar argument as in part (i),
 and the (2) and (3) are from the covariance matrix and $\E[(2)+(3)\mid Y_{i0}]=0$ by a similar argument as in part (ii). Note that for $\rho_Y$, there is Nickell bias for conditional likelihood, but not for the marginal likelihood here.

Combining parts (i)--(iii), every component of the quasi-score $s(\eta)$ has zero expectation under the true DGP, as long as the first two conditional moments are correctly specified. Finally, under Assumptions \ref{assu:model}--\ref{assu:rank} and \ref{assu:est}, the strictly concave quasi-log-likelihood and pointwise LLN yield consistency by the argmax theorem, and a Taylor expansion of the score around the true parameter together with the CLT establishes asymptotic normality.
\hfill\end{proof}

\subsection{Ratio optimality}\label{sec:proof-ratio-optimality}
We adopt Assumptions 3.2--3.6 of \citet{liu2020forecasting}, restated as in our setting as follows.
First, define the slowly diverging sequence as follows.

\begin{definition}[Slowly diverging sequences]\label{def:slowly-diverging-sequences}\(\)
    \begin{enumerate}[label=(\alph*)]
\item $A_N(\pi)=o_{u.\pi}(N^\epsilon)$ for some $\epsilon>0$, if there exists a sequence $\eta_N\to0$ that does not depend on $\pi\in\Pi$ such that $N^{-\epsilon}A_N(\pi)\le\eta_N$.

\item $A_N(\pi)=o(N^+)$, if for every $\epsilon>0$, there exists a sequence $\eta_N(\epsilon)\to0$ such that $N^{-\epsilon}A_N(\pi)\le\eta_N(\epsilon)$.

\item $A_N(\pi)=o_{u.\pi}(N^+)$, if for every $\epsilon>0$, there exists a sequence $\eta_N(\epsilon)\to0$ that does not depend on $\pi\in\Pi$ such that $N^{-\epsilon}A_N(\pi)\le\eta_N(\epsilon)$.
\end{enumerate}
\end{definition}
Intuitively, (a) holds for some $\epsilon$ and uniformly in $\pi$, (b) holds for every $\epsilon$ but only pointwise in $\pi$, and (c) holds for every $\epsilon$ uniformly in $\pi$.

\begin{assumption} [Trimming and bandwidth] \label{assu:trimbandwidth} \(\)
  \begin{enumerate}[label=(\alph*)]
          \item The truncation sequence $C_N$ satisfies $C_N = o(N^{+})$ and $C_N \geq (2\log N)/M_2$. 
          \item The truncation sequence $C_N'$ satisfies $C_N' = C_N + \sqrt{ (2\sigma^2 \log N)/T}$.
          \item The bandwidth sequence $B_N$ is bounded by $\underline{B}_N \le B_N \le \overline{B}_N$, where  $1/\underline{B}_N^2 =o(N^{+})$, $\overline{B}_N(C_N+C_N') = o(1)$, and
          the bounds do not depend on the observed data or $\pi_0 \in \Pi$.  
  \end{enumerate}
\end{assumption}

\begin{assumption}[CRC distribution: tails]\label{assu:tail}
    There exist constants $0<M_1,M_2,M_3,M_4<\infty$ such that for the true distribution $\pi_0\in\Pi$:
    \begin{enumerate}[label=(\alph*)]
      \item $\int_{\|\lambda\|\ge C}\pi_0(\lambda) d\lambda\le M_1e^{-M_2(C-M_3)},$ and
        $\int\|\lambda\|^4\pi_0(\lambda) d\lambda\le M_4.$
      \item $\int_{|y_0|\ge C}\pi_0(y_0) dy_0\le M_1e^{-M_2(C-M_3)},$ and
        $\int y_0^4\pi_0(y_0) dy_0\le M_4.$
    \end{enumerate}
    \end{assumption}
    To estimate the unknown prior nonparametrically, we trim off very large $\lambda_i$ so our kernel estimates do not explode in the tails, but let the trimming threshold $C_N$ grow slowly with $N$.  The exponential tail bound on the prior guarantees little mass beyond $C_N$.  Meanwhile, the kernel bandwidth $B_N$ shrinks just fast enough to capture local features of the prior, but not so fast that variance dominates bias.  Together, these conditions balance trimming and smoothing so the leave-one-out density $\widehat p_{(-i)}$ is consistent.

    \begin{assumption}[CRC distribution: boundedness and smoothness]\label{assu:smooth}
      The conditional density $\pi_0(y_0\mid\lambda)$ is uniformly bounded and
      \[
        \sup_{|y_0|\le C'_N,\,\|\lambda\|\le C_N}\left|\left.\frac{1}{B_N}\int\phi\left(\tfrac{y-y_0}{B_N}\right)\pi_0(y\mid\lambda) dy\right/\pi_0(y_0\mid\lambda)-1\right|=o(1),
      \]
      where sequences $C_N$, $C_N'$, and $B_N$ satisfy Assumption \ref{assu:trimbandwidth}.
      \end{assumption}
We need the conditional density $\pi_0(y_0\mid\lambda)$ to be smooth on the trimmed region, so that convolving it with our Gaussian kernel does not distort its shape substantially.  This prevents spikes or point mass priors on $Y_{i0}\mid\lambda_i$, ensuring the leave-one-out smoothing step yields a valid approximation to the true prior.

    The posterior mean function  and the joint sampling distribution of the sufficient statistic and the initial condition take the form
\begin{align*}
m(\widehat{\lambda},y_0;\pi_0) &= \widehat{\lambda} + \Sigma_V(\theta_0) \frac{\partial  }{\partial \widehat{\lambda} } \log p(\widehat{\lambda},y_0 ; \pi_0), \\
p(\widehat{\lambda},y_0;\pi_0)
&= \int 
\frac{1}{\sqrt{\det(\Sigma_V(\theta_0))}} \phi\left( \Sigma_V(\theta_0)^{-1/2}(\widehat{\lambda} - \lambda)\right) 
\pi_0(\lambda,y_0) d \lambda.
\end{align*}
Also define the following $*$-counterparts by convolving the prior $\pi_0(\lambda,y_0)$ with a Gaussian kernel with bandwidth $B_N$. These $*$-objects are the population targets of the expected leave-one-out kernel estimator
\begin{align*}
  &m_{*}(\widehat{\lambda},y_0;\pi_0,B_N)\\ &= \widehat{\lambda} + \left( \Sigma_V(\theta_0) + B_N^2 I \right) \frac{\partial  }{\partial \widehat{\lambda} }\log p_{*}(\widehat{\lambda},y_0 ; \pi_0,B_N), \\
  &p_{*}(\widehat{\lambda},y_0; \pi_0,B_N) \\
  &= \frac{1}{B_N^d}\int 
  \frac{1}{\sqrt{\det(\Sigma_V(\theta_0) + B_N^2 I)}} \phi\left( (\Sigma_V(\theta_0) + B_N^2 I)^{-1/2}(\widehat{\lambda} - \lambda)\right) 
 \phi\left( \frac{y_{0} - \tilde{y}_{0}}{B_N}\right)  \pi_0(\lambda,\tilde{y}_{0})  d \lambda d \tilde{y}_{0}.
\end{align*}
    
    \begin{assumption}[Posterior mean functions]\label{assu:post-mean} Let $C_N$ be a sequence satisfying Assumption \ref{assu:trimbandwidth}. The posterior mean functions satisfy: 
      \begin{enumerate}[label=(\alph*)]
        \item  $N \displaystyle \iint  \left\|m(\widehat{\lambda},y_{0};\pi_0)\right\|^2 \1\left\{ \left\|m(\widehat{\lambda},y_{0};\pi_0)\right\| \geq C_N \right\} p(\widehat{\lambda},y_0 ; \pi_0) d \widehat{\lambda} d y_0 = o_{u.\pi_0}(N^{+}),$
                \item $ N \displaystyle \iint  \left\|m_{*}(\widehat{\lambda},y_{0};\pi_0,B_N)\right\|^2 \1\left\{ \left\|m_{*}(\widehat{\lambda},y_{0};\pi_0,B_N)\right\| \geq C_N \right\} p(\widehat{\lambda},y_0 ; \pi_0) d \widehat{\lambda} d y_0 = o_{u.\pi_0}(N^{+}),$
                \item $  N \displaystyle \iint  \left\|m(\widehat{\lambda},y_{0};\pi_0)\right\|^2 \1\left\{ \left\|m(\widehat{\lambda},y_{0};\pi_0)\right\| \geq C_N \right\} p_{*}(\widehat{\lambda},y_0 ; \pi_0,B_N) d \widehat{\lambda} d y_0 = o_{u.\pi_0}(N^{+}).$
                \end{enumerate}
    \end{assumption}
    This assumption guarantees that outside a slowly growing ball of radius $C_N$, the contribution to the overall risk is negligible.  In other words, only a vanishing fraction of units have such extreme estimates that they could undermine our uniform risk bound.  We check this not only for the posterior mean $m$ and density $p$, but also for the variance inflated versions $(m^*,p^*)$ that arise from adding the kernel variance $B_N^2$.
    
    \begin{assumption}[Rates for $\widehat\theta$]\label{assu:thetahat}
    The estimator for the common parameters satisfies
    \[
      \E_{\theta_0,\pi_0}\left[\left|\sqrt N(\widehat\rho_Y-\rho_{Y,0})\right|^4\right]=o_{u.\pi_0}(N^+),
      \quad
      \E_{\theta_0,\pi_0}\left[\left|\sqrt N(\widehat\sigma_U^2-\sigma_{U,0}^2)\right|^2\right]=o_{u.\pi_0}(N^+),
    \]
    and similarly for $\rho_{\delta},\sigma_{\varepsilon}^2$.
    \end{assumption}
    Finally, we require our estimator of the common parameters to converge at the usual $\sqrt N$-rate with sufficiently thin tails.  This ensures that plugging $\widehat\theta$ into our empirical Bayes update does not introduce any first-order errors in the risk comparison against the oracle. By Theorem \ref{thm:QMLE}, our QMLE estimator attains the required $\sqrt{N}$-rate and thus fulfills this assumption.

\medskip
    
\noindent\begin{proof}[Proof of Theorem \ref{thm:ratio-optimality}]
In the simple model under Assumption 2.3 (rank condition), the common treatment timing design $W_i(\rho_\delta)$ in \eqref{eq:W_i} is deterministic and satisfies $W_i(\rho_\delta)'W_i(\rho_\delta)$ invertible with finite eigenvalues. Hence, the Moore-Penrose inverse $W_i^{+}(\rho_\delta)=(W_i(\rho_\delta)'W_i(\rho_\delta))^{-1}W_i(\rho_\delta)'$ exists and the sufficient statistic $\widehat\lambda_i(\rho)=W_i^{+}(\rho_\delta)\left(y_{i,1:T}-\widehat\rho_Y y_{i,0:T-1}\right)$ in \eqref{eq:suff-stat} is well defined. Following from \eqref{eq:Y_it}, the covariance of the stacked innovations $\check\Sigma_U(\theta_0)$ is positive semidefinite. Then, the projection noise covariance $\Sigma_{V,i}(\theta_0)=W_i^{+}(\rho_\delta)\check\Sigma_U(\theta_0)\left[W_i^{+}(\rho_\delta)\right]'$ is well defined with finite eigenvalues. 

Since $W_i(\rho_\delta)$ is deterministic and common across $i$ in the simple model, we follow the proof strategy in \citet{liu2020forecasting}, which instead focuses on individual forecasts. Under Assumptions A.1--A.5 governing trimming/bandwidth, CRC tails/smoothness, posterior mean functions, and $\sqrt N$-rates for the common parameters, we obtain the ratio optimality for the jointly estimated individual effects $\alpha_i$ and heterogeneous treatment effects $\delta_{i0}$.
\hfill\end{proof}

\begin{remark}[Extension: rich controls $C_i$]
\normalfont{
Consider the extension in Section \ref{sec:extensions} with a conditional prior $\pi(\lambda_i\mid C_i)$, where
\(
C_i=\left(Y_{i0}, Z_{i,1:T}^{0:J}, X^O_{i,0:T}, X^P_{i0}\right),
\)
$ Z_{i,1:T}^{0:J}$ collects treatment timing and size (w.l.o.g.\ we consider continuous treatment here), $X^O_{i,0:T}$ are strictly exogenous covariate paths, and $X^P_{i0}$ are initial values of predetermined covariates. Now $W_i(\rho_\delta)=W(\rho_\delta,C_i)$ and $\Sigma_{V,i}(\theta)=\Sigma_V(\theta,C_i)$ are functions of $C_i$.

Assume that $W(\rho_{\delta0},C_i)$ has full column rank with the eigenvalues uniformly bounded away from zero over trimmed $C_i$. Following from the continuity of $W(\rho_\delta,C_i)$ in $\rho_\delta$ uniformly over trimmed $C_i$, there exists a compact neighborhood $ \rho_{\delta0}\in\Theta_\rho$ and a constant $0<c<\infty$ such that
\[
\inf_{\rho_\delta\in\Theta_\rho,\,\text{trimmed }C_i}\lambda_{\min}\left(W(\rho_\delta,C_i)'W(\rho_\delta,C_i)\right)\ge c,
\]
so $W^+(\rho_\delta,C_i)$ is well-defined uniformly over $\rho_\delta\in\Theta_\rho$ and trimmed $C_i$. Similarly, the covariance mapping $\Sigma_V(\theta,C_i)$ is smooth in $\theta$ uniformly over a compact neighborhood of $\theta_0$ and trimmed $C_i$. 

With this in place, replace $Y_{i0}$ by $C_i$ throughout Assumptions A.1--A.5. The Tweedie step and the ratio optimality argument then carry over verbatim, now conditional on $C_i$.
}
\end{remark}

\end{document}